\documentclass[10pt,twocolumn]{IEEEtran}
\usepackage{tabularx}
\usepackage[T1]{fontenc}
\usepackage{algorithmic}
\usepackage{graphicx}
\usepackage{amsmath,amssymb,amsfonts,stfloats}
\usepackage{array}
\usepackage{textcomp}
\usepackage[font=small]{caption}
\usepackage{paralist}
\usepackage{lettrine}
\usepackage{cite}
\usepackage{xcolor}
\usepackage{cancel}
\usepackage{mathtools}
\usepackage{booktabs}
\usepackage[caption=false,font=footnotesize]{subfig}
\usepackage{url}
\usepackage{acronym}
\usepackage{multicol}
\usepackage[ruled,vlined]{algorithm2e}

\begin{document}
\title{Cooperative Coherent Multistatic Imaging and Phase Synchronization in Networked Sensing \thanks{This work was partially supported by the European Union under the Italian National Recovery and Resilience Plan (NRRP) of NextGenerationEU, partnership on “Telecommunications of the Future” (PE00000001 - program “RESTART”).}}

\author{Dario~Tagliaferri, Marco~Manzoni, Marouan~Mizmizi, Stefano~Tebaldini, Andrea~Virgilio Monti-Guarnieri, Claudio~Maria~Prati, Umberto~Spagnolini}
\maketitle

\begin{abstract}
Coherent multistatic radio imaging represents a pivotal opportunity for forthcoming wireless networks, which involves distributed nodes cooperating to achieve accurate sensing resolution and robustness. This paper delves into cooperative coherent imaging for vehicular radar networks. Herein, multiple radar-equipped vehicles cooperate to improve collective sensing capabilities and address the fundamental issue of distinguishing weak targets in close proximity to strong ones, a critical challenge for vulnerable road users protection. We prove the significant benefits of cooperative coherent imaging in the considered automotive scenario in terms of both probability of correct detection, evaluated considering several system parameters, as well as resolution capabilities, showcased by a dedicated experimental campaign wherein the collaboration between two vehicles enables the detection of the legs of a pedestrian close to a parked car. Moreover, as \textit{coherent} processing of several sensors' data requires very tight accuracy on clock synchronization and sensor's positioning---referred to as \textit{phase synchronization}---(such that to predict sensor-target distances up to a fraction of the carrier wavelength), we present a general three-step cooperative multistatic phase synchronization procedure, detailing the required information exchange among vehicles in the specific automotive radar context and assessing its feasibility and performance by hybrid Cramér-Rao bound.

\end{abstract}


\section{Introduction}\label{sect:introduction}

Wireless radio sensing in the $3-300$ GHz spectrum is nowadays considered a fundamental task for many novel applications as the catalyst for context awareness (autonomous driving, industrial automation, smart cities) \cite{temdee2018context}.
The effectiveness of an individual sensor within this spectrum is predominantly governed by several key factors, including its hardware configuration, operating frequency, transmitted power, bandwidth, and the specific environmental context in which it operates.
The hardware setup of a sensor is typically influenced by considerations related to manufacturing costs and energy efficiency. Moreover, the choice of operating frequency, transmitted power, and bandwidth is often constrained by regulatory directives (see \cite{rel18,ETSI_TR103593}). In addition, the density of the deployment scenario can significantly impact the sensor's field-of-view (FoV), particularly when comparing dense and sparse environments.
Nevertheless, it is imperative to recognize that a growing number of applications requires a level of environmental awareness beyond the capabilities of individual sensors. In response to this challenge, the prevailing consensus underscores the importance of adopting a \textit{networked sensing} approach. This approach entails the collaboration of multiple sensing entities at various levels to enhance the performance and reliability of the sensing process \cite{liu2022networked}.

The first feature that characterizes a networked sensing system is the figure(s) of merit that is (are) optimized through cooperation. Localization aims at the estimation of the position, velocity, and orientation of a set of targets, and it has been consolidated through years of research~\cite{10287134}. Here, Cramér-Rao lower bound (CRLB) is the preferred performance metric, and it is usually evaluated given specific signal and noise models and assuming a sparse scene, where targets of interest are not mutually coupled in terms of parameters' estimation \cite{Chetty2022_CRB}. Differently, \textit{radio imaging} aims at producing a 2D or 3D map of the complex reflectivity of the environment, to infer both the number of targets and their shape. Imaging quality is not evaluated in terms of CRLB but rather by \textit{resolution}, peak-to-sidelobe ratio (PSLR), and other metrics \cite{cumming}. The localization process, i.e., the estimation of position/velocity/orientation, follows after detection from the radio image. Another important feature of a networked sensing system is the degree of cooperation. Cooperation can be loose, namely sensors perform independent monostatic measurements (i.e., where the transmitter (Tx) is co-located to the receiver (Rx) and share the same hardware) to be mutually exchanged for subsequent fusion. Conversely, cooperation can be tight, i.e., the wireless network operates as a distributed sensor in the acquisition phase, possibly including bistatic measurements (Tx and Rx not co-located). Moreover, as far as pass-band signals are employed, we distinguish between \textit{coherent} and \textit{incoherent} networked sensing, respectively, with or without retaining the carrier phase information in the measured data during the fusion.

%

Networked sensing has been originally explored for distributed radar networks \cite{5703085} or cooperative localization \cite{4802193}, with the clear aim of improving sensing performance. Recent advances in networked sensing, instead, are targeted to integrated sensing and communication (ISAC) systems, where the goal is to fuse sensing with communication over a single physical layer \cite{Liu_survey}. For instance, the work \cite{Caire2022_InfTh_JCS} explores the cooperation possibilities in networked ISAC systems using information theory. The work in \cite{Shi2022_DEVICEFREE} considers the problem of fusing multiple monostatic sensing measurements of the same target from different ISAC nodes to reduce the issue of ghost targets. Information theory is used in \cite{Caire2022_InfTh_JCS} to stress the potential of networked ISAC systems with multiple receiving terminals, while \cite{10049300} discusses the dissemination of sensing signals across different nodes of the network to accomplish cooperative sensing, exploiting epidemic theory. An interesting application of networked sensing is for cell-free multiple-input-multiple-output (MIMO) systems, where a dense network of access points is deployed. In this setting, the benefit of networked sensing is evident \cite{demirhan2023cellfree}. 

Most of the aforementioned works, especially the recent ones on ISAC, consider incoherent networked localization from multiple monostatic acquisitions, where cooperation is enforced to improve the accuracy of parameters' estimation. Multistatic acquisitions (monostatic plus bistatic), however, open novel sensing opportunities \cite{Wymeersch2022} as well as the coherent fusion of measurements, that allows achieving the utter localization performance \cite{wymeersch2023fundamental}. Notice that the localization functionality follows after imaging of the environment and the inference of the number of targets via detection, thus there is an increasing interest in radio imaging with communication networks \cite{5374407,9625204}, that is therefore the topic of this work.  

It is important to underline that sensing data processing over a wireless network poses the practical challenge of \textit{synchronizing} the devices, namely \textit{(i)} estimating and compensating the timing offsets occurring in bistatic acquisitions, arising from clocks' phase and frequency mismatch (clock synchronization) and \textit{(ii)} estimating the positions of individual sensors, necessary to properly fuse the data over a common spatial reference system (spatial synchronization). Synchronization errors cause a space-varying sensor-to-target time-of-fight (TOF) error that compromises the sensing quality and needs to be compensated \cite{10254560}. Notice that, for single bistatic pairs or incoherent data fusion, temporal and spatial synchronization shall guarantee that the TOF error falls within the temporal resolution of single sensors ($\simeq 1/B$, with $B$ being the employed bandwidth), i.e., the error on targets' location ranges from tens to few centimeters. The temporal synchronization accuracy can be achieved with distributed approaches \cite{1425651} or with an accurate common timing reference \cite{Prager2020WirelessSR}, while the required spatial synchronization can be obtained with advanced positioning technologies \cite{7496920}. Still, the resolution of the final radio image does not significantly improve with incoherent data fusion, although localization may benefit for an increased signal-to-noise ratio (SNR). 

Differently, \textit{coherent} data fusion allows drastically improve the radio imaging resolution (as well as localization accuracy) but the synchronization requirements (i.e., clock and positioning requirements) are much tighter. In particular, we refer to \textit{phase synchronization} as the capability of each sensor (and ultimately of the network of sensors) to predict the the TOF to within an accuracy of a small fraction of the carrier frequency ($\simeq 1/f_0$, with $f_0\gg B$). We stress that, even for perfect clock synchronization, e.g., by disciplined oscillators~\cite{9848428}, \textit{millimeter-level} sensors' positioning is needed for $f_0 > 30$ GHz, thus only data-driven approaches can be used. Quite interestingly, phase synchronization among different terminals for coherent imaging has been studied in depth in the field of space-borne and air-borne synthetic aperture radar interferometry (InSAR) and tomography (TomoSAR), to the aim of tracking millimetric deformations and provide high-resolution 3D imaging \cite{ferretti01,berardino02,Phase_cal_TGRS,6557505,5299035,Tebaldini2023}. Still, no specific investigation has ever been done for coherent networked imaging in urban scenarios, where the topology of the employed wireless network, the potential number of nodes, their capabilities, as well as the nature of targets of interest are generally different. Moreover, the overall imaging quality in satellite-based remote sensing systems is somewhat under control, as choosing the satellites' orbits to maximize the imaging resolution is a viable option \cite{Tebaldini2017_tessellation}. As a last difference, imaging in remote sensing missions is usually performed offline, while cellular or vehicular networks require an online information exchange to produce high-resolution images via cooperation.

\begin{figure}[!b]
    \centering
    \includegraphics[width=\columnwidth]{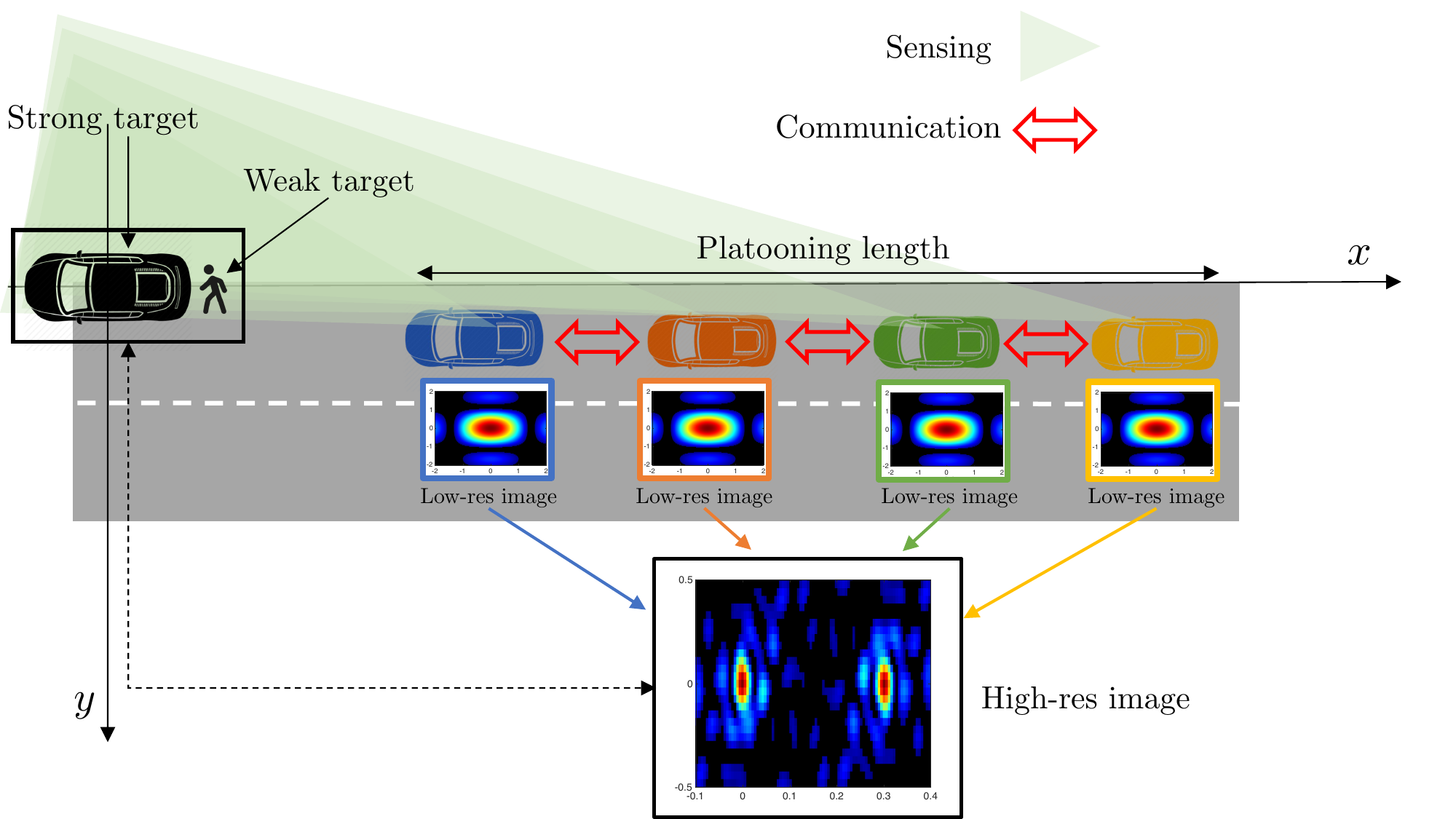}
    \caption{Vehicular scenario considered in the paper. A weak target (the pedestrian) is standing very close to a strong target (the vehicle). It is important to detect pedestrians to monitor their status for safety-critical applications.}
    \label{fig:scenario}
\end{figure}

\textit{Contribution}: To avoid excessive abstract settings, this paper focuses on vehicular wireless networks for autonomous driving and vulnerable road user (VRU) protection \cite{9787407}. Nowadays commercialized cars mount a plethora of sensors (cameras, lidars, and radar) ~\cite{Marti2019ADAS_sensors} to effectively sense and perceive the surrounding environment.
In particular, high-resolution and imaging radars in W-band ($76-81$ GHz) are under increasing investigation for their comparatively low cost, weather robustness, and potential good resolution (1 deg in azimuth, tens of cm in range)~\cite{8830483}. Unfortunately, the available bandwidth of commercial automotive radars decreases with the sensing range, and when multiple radars simultaneously operate over the same area, mutual interference requires further limiting the available bandwidth, with a drastic reduction of imaging capabilities \cite{ETSI_TR103593}. Although the research on high-end radars is active \cite{Waldschmidt2021_future_radar}, the manufacturing cost pushes for the usage of multiple platforms to increase the resolution. The most relevant work in this direction is \cite{9318740}, where a vehicle is equipped with two cabled radars that cooperate to obtain a coherent image of the environment. 
However, as far as vehicles are equipped with a vehicle-to-vehicle (V2V) communication interface~\cite{ETSI_TS103300-3}, unprecedented high-resolution imaging can be accomplished by coherent multistatic networked sensing over multiple vehicles. Fig. \ref{fig:scenario} shows the considered use case. Several vehicles are traveling on the same lane in a platooning configuration. Each vehicle is equipped with a low-end, off-the-shelf automotive radar working in the $76-81$ GHz band, that operates in frequency division multiplexing (FDM) w.r.t. others, and allows producing an image of the environment at a resolution dictated by the hardware capabilities of the single devices. A target of interest, i.e., a pedestrian approaching the lane, is located near a highly reflective target, e.g., a parked car. The resolution of the single radar is for not enough to distinguish the two targets, thus the goal of the vehicular network is to enforce cooperation to obtain a high-resolution image from the coherent fusion of single images of the intended area to detect and discriminate the pedestrian from the parked car.

The contributions of the paper are as follows:

\begin{itemize}
 
    \item We formalize the cooperative multistatic phase synchronization problem, addressing both the clocks' phase and frequency mismatch as well as and sensors' positioning. We outline a data-based cooperative procedure that leverages a set of calibration targets in the environment as well as a peculiar information exchange over the vehicular network. The feasibility and requirements of phase synchronization are then assessed by hybrid Cramér-Rao bound (HCRB) on estimated phases at each sensor, for random vehicular network topology, devising sufficient conditions for the solution of the phase synchronization problem as well as practical guidelines for the selection of calibration targets given the number of vehicles. Noticeably, the discussed cooperative multistatic synchronization method is general and can be applied, with due adaptations, to \textit{any} networked sensing system.

    \item We demonstrate the significant benefits of coherent multistatic imaging in the aforementioned vehicular scenario, in terms of both resolution and corresponding probability of correct detection (PCD) of a weak target near to a highly reflective one. We show, with experimental automotive radar data, that a vehicle equipped with an off-the-shelf low-end radar ($180$ MHz bandwidth at $77$ GHz carrier frequency) can aim to produce centimeter-level accurate images of the surroundings via cooperation with other vehicles. Remarkably, we provide an experimental demonstration of the capability of accurately detecting a pedestrian standing at $30$ cm from a parked car enforcing a multistatic cooperation between two vehicles at $5$ m distance one to the other, whereby single radars cannot discern the two targets.

\end{itemize}



\textit{Organization}: The remainder of the paper is organized as follows: Section \ref{sect:system_model} outlines the system model, Section \ref{sect:_synchro_and_pos} discusses the multistatic cooperative phase synchronization, Section \ref{sect:HCRB} quantifies the synchronization performance via HCRB, Section \ref{sect:PCD} evaluated the PCD in a generic multistatic setup. with and without residual phase errors, Section \ref{sec:real_data} presents the experimental high-resolution imaging demonstration while Section \ref{sec:conclusions} concludes the paper. Appendix \ref{sect:FEDT} recalls the background on diffraction tomography, that allows quantifying the resolution of any radio imaging experiment. 

\textit{Notation}: The following notation is adopted in the paper: bold upper- and lower-case letters describe matrices and column vectors, respectively. $\mathbf{I}$ and $\mathbf{1}$ denote the identity matrix and a vector of all ones whose sizes are specified case-by-case. Matrix transposition is indicated as $\mathbf{A}^T$. $\mathrm{trace}(\mathbf{A})$ denotes the trace of $\mathbf{A}$. Vector norm is $\|\mathbf{a}\|$. $|\mathcal{A}|$ denotes the cardinality of set $\mathcal{A}$. With  $\mathbf{a}\sim\mathcal{CN}(\boldsymbol{\mu},\mathbf{C})$ we denote a multi-variate circularly complex Gaussian random variable $\mathbf{a}$ with mean $\boldsymbol{\mu}$ and covariance $\mathbf{C}$. $\mathbb{R}$ and $\mathbb{C}$ stand for the set of real and complex numbers, respectively. $\delta_{n}$ is the Kronecker delta.

\begin{figure}[!t]
    \centering
    \includegraphics[width=0.8\columnwidth]{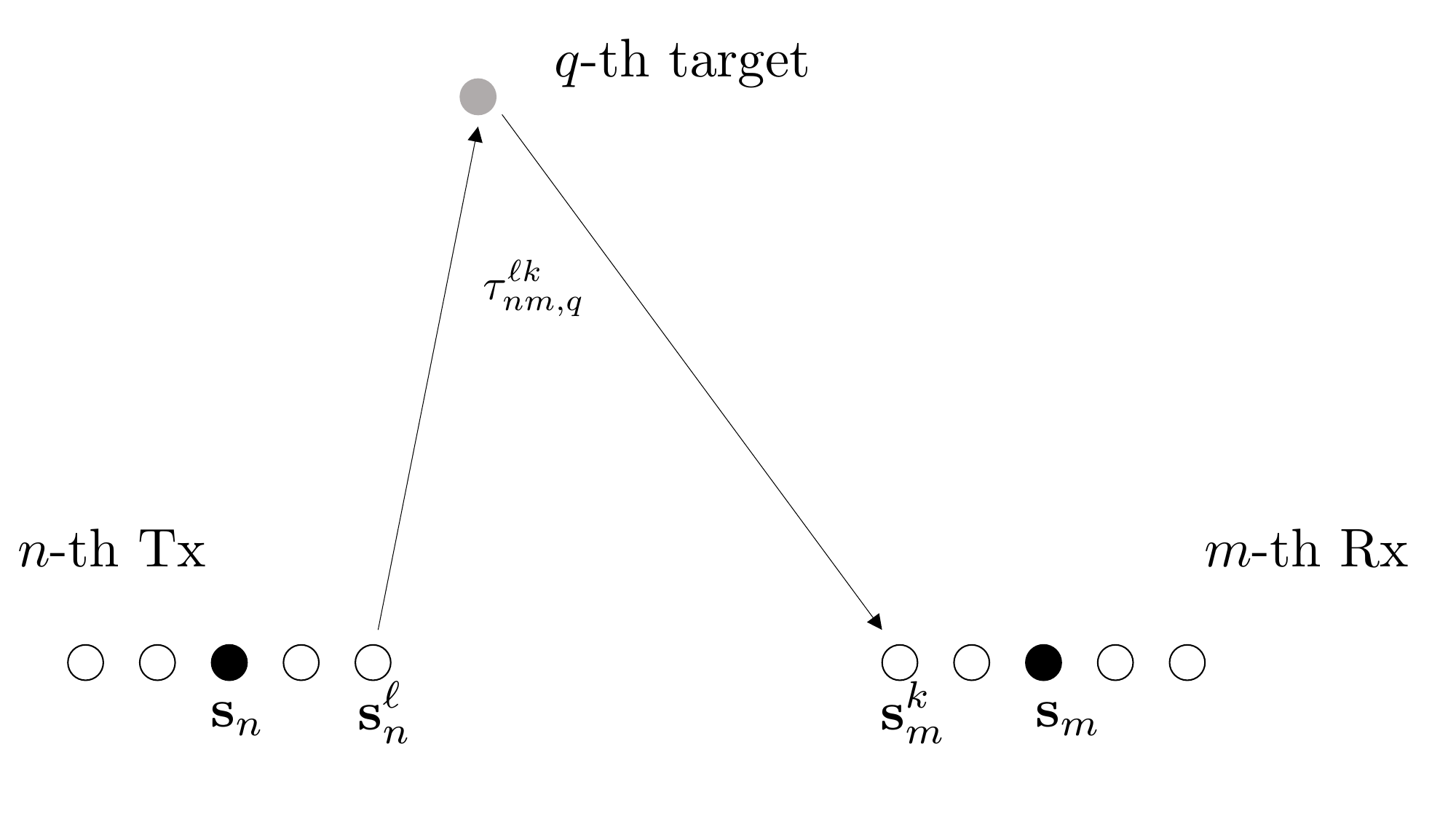}
    \caption{Reference system and notation.}
    \label{fig:ref_system}
\end{figure}

\section{System Model}\label{sect:system_model}

Let us consider the network of $N$ vehicles shown in Fig. \ref{fig:scenario}. Each vehicle is equipped with a radar sensor that is located at position $\mathbf{s}_n \in \mathbb{R}^{2 \times 1}$ and is used in both monostatic and bistatic operation to image the environment, composed of $Q$ targets at positions $\mathbf{x}_q \in \mathbb{R}^{2 \times 1}$, for $q=1,...,Q$. Herein, we assume a global reference system where each sensor knows its position and orientation with a certain accuracy, e.g., by GPS or other technology. Each sensor implements a spatial aperture $\mathcal{A}_n$, either physical or synthetic, where the location of the $\ell$-th antenna of the $n$-th sensor is $\mathbf{s}^{\ell}_n$. The sensors operate in FDM, namely the $n$-th radar operates on a different spectrum portion $B$, centered around $f_n$. The sketch of the reference system and notation is in Fig. \ref{fig:ref_system}

The model of the signal received at the $k$-th antenna of the $m$-th Rx, due to the transmission from the $\ell$-th antenna of the $n$-th Tx sensor is reported in \eqref{eq:Rxsignal_time_demod},
\begin{figure*}[!t]
\begin{equation}\label{eq:Rxsignal_time_demod}
\begin{split}
    d_{nm}^{\ell k}(t) &= \sum_{q=1}^Q \varrho_{q} e^{-j \theta_q} \, g_{\mathrm{Tx}}\left(t-\tau_{nm,q}^{\ell k} + \varepsilon_n(t-\tau_{nm,q}^{\ell k}) - \varepsilon_m(t)\right) e^{j 2 \pi f_n \left(t-\tau_{nm,q}^{\ell k} + \varepsilon_n(t-\tau_{nm,q}^{\ell k})- \varepsilon_m(t)\right)} + w_{nm}^{\ell k}(t)\\
    & = \sum_{q=1}^Q \varrho_{q}e^{-j \theta_q}\, g_\mathrm{Tx}\left((1+\beta_{nm})t- (1 + \beta_{n}) \tau_{nm,q}^{\ell k} + \kappa_{nm}\right) \, e^{j \left[2 \pi f_n (1+\beta_{nm}) t - 2 \pi f_n (1 + \beta_n) \tau_{nm,q}^{\ell k} + \alpha_{nm}\right]} +  w_{nm}^{\ell k}(t)
\end{split}
\end{equation}
\hrulefill
\end{figure*}
where 
\begin{itemize}
    \item $\varrho_{q}$ is the scattering magnitude of the $q$-th target, depending on its radar cross section (RCS) of the target and assumed equal for all the Tx-Rx pairs for simplicity;
    \item $\theta_q$ is the target's backscattering phase, related to its physical composition and shape, which is assumed to be the same for each Tx-Rx pair. This latter holds for perfectly isotropic targets, or small small-angle views, as considered in Fig. \ref{fig:scenario} and experimentally observed in~\cite{rs15030845}; 
    \item $g_\mathrm{Tx}\left(t\right)$ is the base-band Tx waveform of energy $E_g$, bandwidth $B$ and duration $T$ (equal for all sensors and antennas, for simplicity of exposition);
    \item $\tau_{nm,q}^{\ell k}$ is the TOF for the $q$-th target, defined as
    \begin{equation}\label{eq:delays}
        \tau_{nm,q}^{\ell k} =  \frac{\| \mathbf{x}_q - \mathbf{s}^\ell_n\| + \| \mathbf{s}^k_m - \mathbf{x}_q\| }{c},
    \end{equation}
    where $c$ is the wave speed;
    \item $ \varepsilon_n(t)$ denotes the clocks' timing mismatch between Tx and Rx, that within short pulse duration $T$ is modeled as~\cite{Vannicola83}
    \begin{equation}\label{eq:timing_error}
    \varepsilon_n(t) = \kappa_n + \beta_n t,
    \end{equation}
    with $\kappa_n$ and $\beta_n$ being the time and frequency offsets (the latter normalized to $f_n$) at the $n$-th sensor, respectively;
    \item $w_{nm}^{\ell k}(t)\sim \mathcal{CN}(0, \sigma_w^2\delta_{\ell-k}\delta_{n-m})$ is the additive noise corrupting the signal.
\end{itemize}
In \eqref{eq:Rxsignal_time_demod}, the clocks' timing mismatch transforms into a TOF shift $\kappa_{nm}=\kappa_{n}-\kappa_{m}$ on the baseband waveform, and a phase mismatch $\alpha_{nm} = \alpha_n - \alpha_m = 2 \pi f_n\kappa_{nm}$ that affects the carrier. The relative Tx-Rx frequency mismatch $\beta_{nm} = \beta_m-\beta_n$, instead, leads to a time dilatation/contraction (for $\beta_{nm}>0$ or $\beta_{nm}<0$) on both the base-band waveform and the carrier. In monostatic sensing setups, we have $\kappa_{nn}=\alpha_{nn}=0$ and $\beta_{nn}=0$, as Tx and Rx share the same clock. 

The Rx signal \eqref{eq:Rxsignal_time_demod} in bistatic setups ($\kappa_{nm}\neq 0$, $\alpha_{nm}\neq 0$, $\beta_{nm}\neq 0$) distorts \textit{range compression} (i.e., demodulation at nominal frequency $f_n$ and matched filtering), for frequency mismatch $\beta_{nm}$. The Rx signal after demodulation by $f_n$ would be affected by a residual term $e^{j 2 \pi f_n \beta_{nm} t}$, whose period might be shorter than the pulse duration $T$ in typical conditions, i.e., $1/(f_n \beta_{nm}) < T$. Thus, the useful signal power after matched filtering can possibly vanish. To obviate this latter issue, it is necessary to estimate the normalized frequency offset $\beta_{nm}$ over a short period of time, typically comparable with common pulse duration (tens of $\mu$s) such that model \eqref{eq:timing_error} holds. For radar systems operating at millimeter waves and pulse duration of tens of $\mu$s, the accuracy of the estimate $\widehat{\beta}_{nm}$ is usually enough to neglect any time-varying phase term after demodulation, as discussed in detail in Section \ref{subsect:freq_est}.

After compensating for the frequency mismatch by $e^{-j 2 \pi f_n \widehat{\beta}_{nm} t}$, and matched-filtering the data, one gets \eqref{eq:Rxsignal_time_RC},
\begin{figure*}[t!]
\begin{equation}\label{eq:Rxsignal_time_RC}
\begin{split}
     y_{nm}^{\ell k}(t) &= \left[d_{nm}^{\ell k}(t)\, e^{-j 2 \pi f_n(1+\widehat{\beta}_{nm}) t}\right]  * g^*_\mathrm{Tx}(-t) \simeq  \sum_{q=1}^Q  \varrho_{q}\, g\left(t - (1 + \beta_{n}) \tau_{nm,q}^{\ell k} + \kappa_{nm}\right) e^{-j \varphi_{nm,q}^{\ell k}} + n^{\ell k}_{nm}(t) 
\end{split}
\end{equation}
\hrulefill
\end{figure*}
where $g(t)$ is the range-compressed pulse and 
\begin{equation}\label{eq: phase model}
    \varphi_{nm,q}^{\ell k} = 2 \pi f_n (1 + \beta_n) \tau_{nm,q}^{\ell k} - \alpha_{nm} + \theta_q
\end{equation}
is the residual phase term. Noise after range compression is $n^{\ell k}_{nm}(t)$. Notice that Rx signal \eqref{eq:Rxsignal_time_RC} experiences the apparent TOF due to the Tx frequency drift $\beta_n$ in both bistatic and monostatic setups.  

\subsection{Rx signal in the wavenumber domain}

We can review the Rx signal \eqref{eq:Rxsignal_time_RC} over the ensemble of antennas to capture the imaging capability in the wavenumber domain. We assume, for simplicity, perfect synchronization ($\kappa_{nm}=0, \alpha_{nm}=0, \beta_n=0$) and a frequency-flat range-compressed pulse $g(t)$ over $B$, thus $|G(f)| = \sqrt{E_g/B}$. By first operating a Fourier transform, and then following the procedure outline in Appendix \ref{sect:FEDT}, we obtain: 
\begin{equation}\label{eq:Rxsignal_wavenumbers}
\begin{split}
     y_{nm} (\mathbf{k}) \hspace{-0.1cm} = \hspace{-0.1cm}   \sum_{q=1}^Q  \eta_{q} \, e^{-j[\mathbf{k}^T \mathbf{x}_q+\theta_q]} \, \Gamma_{nm,q}(\mathbf{k}) + z_{nm}(\mathbf{k})
\end{split}
\end{equation}
where \textit{(i)} $|\eta_{q}|  = \varrho_q \sqrt{E_g/B}$ is the scattering amplitude of the $q$-th target including path-loss $\varrho_q$, \textit{(ii)} $\mathbf{k}$ is the illuminated wavenumber (see Appendix \ref{sect:FEDT}), while \textit{(iv)} 
\begin{equation}\label{eq:gamma}
    \Gamma_{nm,q}(\mathbf{k}) = \begin{dcases}
    1 & \text{if} \,\,\mathbf{k} \in \mathcal{K}_{nm}(\mathbf{x}_q)\\
    0 & \text{elsewhere}
    \end{dcases}
\end{equation}
is the FT of the $q$-th target's reflectivity function sampled in the set of observed wavenumbers by the $nm$-th Tx-Rx pair
\begin{equation}\label{eq:wavenumber_coverage_nm}
    \mathcal{K}_{nm}(\mathbf{x}_q) = \bigcup_{\ell,k} \mathcal{K}_{B}(\mathbf{s}^\ell_n, \mathbf{s}^k_m, \mathbf{x}_q),
\end{equation}  
whose widths along $k_x$ and $k_y$ dictate the $x$ and $y$ resolution of the $nm$-th image for the $q$-th target.
Wavenumber coverage for the $nm$-th Tx-Rx pair is calculated from \eqref{eq:wavenumber_coverage_two_arrays} of Appendix \ref{sect:FEDT}, by plugging $\mathbf{s}_\mathrm{Tx}=\mathbf{s}^\ell_{n}$, $\mathbf{s}_\mathrm{Rx}=\mathbf{s}^k_{m}$\footnote{Propagation phase $e^{-j[\mathbf{k}^T\mathbf{x}_q+\theta_q]}$ in \eqref{eq:Rxsignal_wavenumbers} accounts for the position of the target in the global coordinate system, rather than of its relative position w.r.t. Tx and Rx sensors. Phase in \eqref{eq:Rxsignal_wavenumbers} coincides with \eqref{eq: phase model} by properly accounting for the propagation from the origin of the global coordinate system to Tx and Rx phase centers. However, these latter contributions are neglected for simplicity of exposition (they are however irrelevant in the evaluation of the image resolution, see Appendix \ref{sect:FEDT}).}.
Finally, term $z^{\ell k}_{nm}(\mathbf{k})\sim\mathcal{CN}(0,N_0 \delta_{n-m}\delta_{\ell-k})$ denotes the noise in the wavenumber domain. The overall spectral covered region by \textit{all} the monostatic and bistatic pairs is the union of each tile \eqref{eq:wavenumber_coverage_nm}, yielding
\begin{equation}\label{eq:wavenumber_coverage_overall}
    \mathcal{K}(\mathbf{x}_q) = \bigcup_{n,m} \mathcal{K}_{nm}(\mathbf{x}_q)
\end{equation} 
from which it follows the resolution bound for the coherent summation of all the images. 

%
%
%
%
%
%


\subsection{Image formation}

The image produced by the $nm$-th Tx-Rx pair in the spatial domain is obtained by \textit{back-projection}, i.e., 2D spatial matched filtering. The expression of the back-projection is reported in \eqref{eq:image_sync} \cite{cumming},
\begin{figure*}[!t]
\begin{equation}\label{eq:image_sync}
\begin{split}
      I_{nm}(\mathbf{x}) & = \sum_{k,\ell} y_{nm}^{\ell k}(t=(1 +\beta_{n}) \tau^{\ell k}_{nm}(\mathbf{x}) + \kappa_{nm}) \,e^{+j 2\pi f_{n}\left(1+\beta_{n}\right)\tau^{\ell k}_{nm}(\mathbf{x})}  
       \simeq e^{-j \alpha_{nm}} \sum_{q=1}^Q \eta_{q} e^{-j\theta_q} h_{nm,q}(\mathbf{x}-\mathbf{x}_q) + n_{nm}(\mathbf{x})
\end{split}
\end{equation}
\hrulefill
\end{figure*}
where the image is evaluated on a pre-defined grid of pixels $\{\mathbf{x}\}$ in a \textit{global} coordinate system, common to all the sensors in the network. This is a fundamental assumption that applies to most practical cases and eases the derivations in the paper. The image formation is herein expressed in the time domain from the range-compressed data \eqref{eq:Rxsignal_time_RC}, where $\tau^{\ell k}_{nm}(\mathbf{x})$ is the TOF \eqref{eq:delays} for an arbitrary pixel $\mathbf{x}$. The back-projection evaluates the Rx data in the apparent TOF $(1 + \beta_n) \tau^{\ell k}_{nm}(\mathbf{x}) + \kappa_{nm}$ and compensates for the propagation phase, then it sums over the available measurements. The very same operation can be expressed in the wavenumber domain, but not reported for simplicity of exposition. 
The final image $I_{nm}(\mathbf{x)}$ is a collection of $Q$ impulse response functions $h_{nm,q}(\mathbf{x})$---obtained as the 2D inverse Fourier transform of $\Gamma_{nm,q}(\mathbf{k})$ in \eqref{eq:gamma}--- shifted and  scaled, while $n_{nm}(\mathbf{x})$ is the noise in the spatial domain, generally not white. 

It is worth highlighting that the single image \eqref{eq:image_sync} is formed under the perfect knowledge of the actual Tx frequency offset $\beta_n$, the TOF $\tau^{\ell k}_{nm}(\mathbf{x})$ and the timing offset $\kappa_{nm}$. If all the aforementioned parameters were perfectly known, the overall image exploiting all the Tx-Rx pairs would be formed as
\begin{equation}\label{eq:image_sync_tot}
    I(\mathbf{x}) = \sum_{n,m}  I_{nm}(\mathbf{x}) e^{j\alpha_{nm}}. 
\end{equation}
where the residual phase $\alpha_{nm}$ shall be properly compensated, as detailed in Section \ref{subsect:fine_sync}.
%
%

\begin{figure}[!t]
    \centering
     \includegraphics[width=\columnwidth]{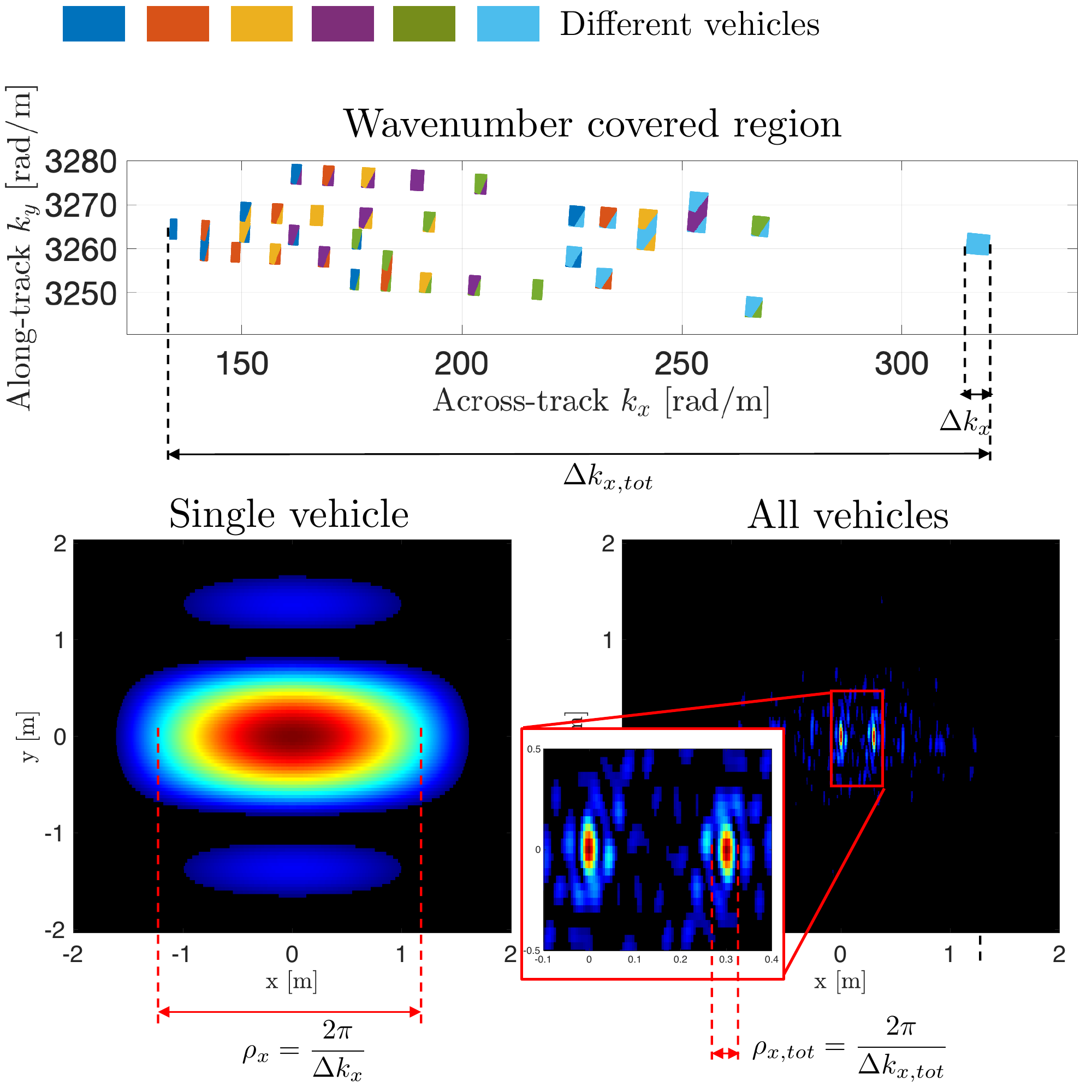}\label{fig:toy_example}
    \caption{Example of imaging of two closely spaced point targets by a network of $6$ cooperating vehicles: (top) covered region in the wavenumber domain (different colors indicate different vehicles, wavenumber tiles with a single color shows monostatic acquisitions, otherwise bistatic ones); (bottom left) image produced by a single vehicle; (bottom right) image produced by coherently fusing the $36$ multistatic images. }
    \label{fig:toy_example}
\end{figure}

\subsection{Imaging example in a simple scenario}

Making reference to the vehicular network scenario portrayed in Fig. \ref{fig:scenario}, Fig. \ref{fig:toy_example} shows an imaging example when 6 vehicles, randomly deployed on the same lane over a span of $60$ m along $x$ (along-track), aim at imaging two closely spaced targets located in the origin of the coordinate system, at a minimum distance of $20$ m along $x$ (maximum distance $80$ m) and $3$ m along $y$ (across-track). Each vehicle uses $B=100$ MHz bandwidth (as typical for medium-to-long range automotive radars \cite{ETSI_TR103593}) and implements an aperture of $20$ cm along $x$. Fig. \ref{fig:toy_example}(top) shows the wavenumber coverage $\mathcal{K}(\mathbf{0})$ of the multistatic setup where single tiles $\mathcal{K}_{nm}(\mathbf{0})$ are portrayed in different colors. Fig. \ref{fig:toy_example}(bottom) shows the corresponding images produced by a single vehicle or by the whole vehicular network, respectively. While the single vehicle cannot distinguish the two targets, due to insufficient resolution, the whole networked sensing system can, due to the coherent combination of multiple wavenumber tiles from different acquisitions. The following Section \ref{sect:_synchro_and_pos} details the practical acquisition of the nominal parameters necessary to coherently combine all the measurements, attaining the imaging quality shown in Fig. \ref{fig:toy_example}.

\section{Cooperative multistatic Synchronization}\label{sect:_synchro_and_pos}

The cooperative multistatic synchronization consists of estimating the required parameters from the Rx data at each sensor to: \textit{(i)} form single images with \eqref{eq:image_sync} and \textit{(ii)} coherently combine the single images as for \eqref{eq:image_sync_tot}. Synchronization errors can be caused by timing offsets $\kappa_{nm}$, frequency offsets $\beta_{n}$ and positioning errors (affecting the time of flight $\tau_{nm,q}$).
Timing and frequency offsets occur due to differences in the hardware oscillators used in Tx and Rx. Positioning errors, when larger than a fraction of the wavelength, cause a space-varying offset in the propagation phase (i.e., a phase screen) that prevents the coherent fusion \eqref{eq:image_sync_tot}. Notice that data-driven bistatic synchronization has been addressed in other works, e.g., see \cite{10254560}, to enable unbiased localization of targets or the formation of single images. For coherent image fusion \eqref{eq:image_sync_tot}, however, the positioning requirements are much tighter and require a different approach \cite{Phase_cal_TGRS}.

The proposed synchronization procedure consists of three main steps: \textit{(i)} estimation of the frequency offsets $\{\beta_n\}$, \textit{(ii)} coarse synchronization, namely estimation of sensors' positions $\{\mathbf{s}^\ell_n\}$ and timing offsets $\{\kappa_{nm}\}$, leading to a coarse knowledge of the TOFs $\{\tau_{nm}^{\ell k}(\mathbf{x})\}$ to form the single images $I_{nm}(\mathbf{x})$ and \textit{(iii)} fine synchronization by estimation of $\{\alpha_{nm}\}$ and refined estimation of $\{\tau_{nm}^{\ell k}(\mathbf{x})\}$. In the following, $\widehat{\cdot}$ is used to indicate quantities that are available prior to the synchronization procedure, $\overline{\cdot}$ denotes quantities after steps \textit{(i)} and \textit{(ii)}, while $\widetilde{\cdot}$ denotes parameters after step \textit{(iii)}.

\textbf{Cooperation requirements}: Before delving into multistatic phase synchronization, it is worth underlining the basic cooperation requirements for the wireless network. We assume that each sensor has an a-priori estimate of its position in global coordinates, $\{\widehat{\mathbf{s}}_n^\ell\}$ $\forall n, \forall \ell$, e.g., by GPS or other technology, but the accuracy is insufficient to proceed to form the multistatic images and their coherent summation, as positioning accuracy is required to be within a fraction of the wavelength. Moreover, pursuing a data-based synchronization procedure requires to have a coarse a-priori temporal synchronization of Tx pulses, such that $\kappa_{nm}$ is bounded to be of the same order of magnitude the duration of the compressed pulse $g(t)$, i.e., $|\kappa_{nm}| \sim 1/B$. This is the common background for distributed synchronization \cite{1425651}, still not sufficient for the tight requirements of coherent networked imaging. 
From the communication perspective, the cooperative phase synchronization procedure presupposes an exchange among sensors of $\widehat{\mathbf{s}}^\ell_n$, that for linear arrays we can parameterize with the position of the phase center $\widehat{\mathbf{s}}_n$ and the orientation of the array $\widehat{\Psi}_n$. Furthermore, after the frequency offset estimation of Section \ref{subsect:freq_est}, that is performed without any exchange of information, the $n$-th sensor shall exchange its monostatic image $I_{nn}(\mathbf{x})$ to all the others in the network. Monostatic images play a pivotal role in both coarse and fine synchronization steps, as described in the following. 

%

\subsection{Frequency Offset Estimation} \label{subsect:freq_est}

According to model \eqref{eq:image_sync}, the formation of images $I_{nm}(\mathbf{x})$ requires an estimate of $\beta_{n}$. The estimation of $\{\beta_n\}$ starts from the $N-1$ available estimates of differential frequency offsets $\{\widehat{\beta}_{nm}\}$ at every sensor, that is needed to enable range compression (Section \ref{sect:system_model}). The accuracy of $\beta_{nm}$ is upper-bounded by $\sigma_\beta \leq 1/(f_n T)$, i.e., inversely proportional to the duration of a single pulse. For a typical duration $T$ in the order of $50-100$ $\mu$s and $f_n=77$ GHz, the dispersion of the propagation phase due to an error over the estimation of $\beta_{nm}$ is $\sigma_\varphi = 2 \pi f_n \sigma_\beta \tau \simeq1$ deg, at the reference distance of $50$ m. Such a figure is equivalent to the dispersion of phase noise at SNR $\simeq 30$ dB and it allows guarantee the neglect of the frequency estimation error in the following synchronization procedure.
Therefore, in the considered multistatic setup, we assume that the $m$-th Rx terminal has $N-1$ available accurate estimates $\widehat{\beta}_{nm}$ of frequency mismatches, for $n=1,...N$, $n\neq m$ from which to obtain the estimates of $N$ absolute frequency offsets. The linear least squares problem admits a solution except a constant term $\delta\beta$ common to all the estimates: 
\begin{equation}\label{eq:est_beta}
    \overline{\beta}_n = \beta_n + \delta\beta.
\end{equation}
Notice that, after compensating for $\overline{\beta}_n$, the residual frequency offset $(1+\delta\beta)$ will generate a compression/dilation in the formed images, that amounts to a few millimeters for oscillators with few to tens parts-per-million of frequency error.

\subsection{Coarse Synchronization}\label{subsect:coarse_sync}

Once the frequency offset estimation is available, sensors proceed to refine the a-priori available positions and orientations, i.e., $\{\widehat{\mathbf{s}}^\ell_n\}$, with the \textit{coregistration} procedure \cite{4215064}, to form both monostatic and bistatic images. Subsequently, timing offsets $\kappa_{nm}$ are estimated from the previously generated images to correct bistatic images such that to have a set of multistatic images $\{I_{nm}(\mathbf{x})\}$ coarsely aligned to a \textit{common spatial grid} to a fraction of the spatial resolution of single images. The coregistration leverages on monostatic intensity images $|I_{nn}(\mathbf{x})|$ and consists of finding the rigid-body roto-translation between a master monostatic image $|I_{\mathrm{ref}}(\mathbf{x})|$ and a slave one $|I_{nn}(\mathbf{x})|$, both evaluated (focused) over a \textit{common spatial grid}. Therefore: 
\begin{equation}\label{eq:coregistration}
    [\widehat{\Psi}_{n}, \widehat{\boldsymbol{\Delta}}_{n}] = \underset{\Psi, \boldsymbol{\Delta}}{\text{argmax}} \left\{ \sum_{\mathbf{x}} \big\lvert I_{\mathrm{ref}}(\mathbf{x})\big\rvert \big\lvert I_{nn}(\mathbf{R}_{\Psi}\mathbf{x}-\boldsymbol{\Delta})\big\rvert\right\}
\end{equation}
where the summation is over the pixels, $\mathbf{R}_{\Psi} \in \mathbb{R}^{2 \times 2}$ is the rotation matrix by angle $\Psi$ and $\boldsymbol{\Delta}$ is the translation offset. The corrected sensors' positions are
\begin{equation}\label{eq:refinedPosition}
    \overline{\mathbf{s}}^\ell_n = \mathbf{R}_{{\widehat{\Psi}}_{n}}\left(\widehat{\mathbf{s}}^\ell_n - \widehat{\mathbf{s}}_n \right)-\widehat{\boldsymbol{\Delta}}_n + \widehat{\mathbf{s}}_n.
\end{equation}
and are used to form both monostatic and bistatic images with \eqref{eq:image_sync}, by substituting \eqref{eq:refinedPosition} into the generic space-varying TOF $\tau_{nm}^{\ell k}(\mathbf{x})$ and by letting $\kappa_{nm}=0$. However, the TOF bias due to $\kappa_{nm} \neq 0$ in bistatic data maps into an image deformation that must be corrected \cite{10254560}. Therefore, we can consider to use the brightest target coincident in both the reference monostatic image $|I_{\mathrm{ref}}(\mathbf{x})|$ and the $nm$-th bistatic one $|I_{nm}(\mathbf{x})|$, i.e., 
\begin{equation}
    \mathbf{x}^{\mathrm{max}}_{\mathrm{ref}} = \underset{\mathbf{x}}{\text{argmax}} \,|I_{\mathrm{ref}}(\mathbf{x})|, \,\,\, \mathbf{x}^{\mathrm{max}}_{nm} = \underset{\mathbf{x}}{\text{argmax}} \, |I_{nm}(\mathbf{x})|,
\end{equation}
to estimate the timing offset as
\begin{equation}\label{eq:kappa_est}
    \overline{\kappa}_{nm} = (1 + \overline{\beta}_n)^{-1} \left[ \overline{\tau}_{nm}( \mathbf{x}^{\mathrm{max}}_{nm}) - \overline{\tau}_{nm}( \mathbf{x}^{\mathrm{max}}_{\mathrm{ref}}) \right]
\end{equation}
where
\begin{equation}
    \overline{\tau}_{nm}( \mathbf{x}) = \frac{\| \mathbf{x} - \overline{\mathbf{s}}_n\| + \| \overline{\mathbf{s}}_m - \mathbf{x}\| }{c}
\end{equation}
is the parametric TOF using the estimated positions of the phase centers of Tx and Rx. Now, bistatic images are correctly formed with \eqref{eq:image_sync}, properly accounting for the timing offset. The formed multistatic images $\{I_{nm}(\mathbf{x})\}$ are coarsely aligned on the common spatial grid, thus a further fine synchronization step is required for coherent fusion \eqref{eq:image_sync_tot}.

\subsection{Fine Synchronization}\label{subsect:fine_sync}


The fine synchronization procedure operates on formed images $\{I_{nm}(\mathbf{x})\}$ after the coarse synchronization, and it has the goal of removing all phase terms due to residual positioning and clock synchronization errors. At the end of fine phase synchronization, it is possible to coherently combine the phase-calibrated images
\begin{equation}
I_{nm}^{\mathrm{cal}}(\mathbf{x})=I_{nm}(\mathbf{x})e^{j\varphi_{nm}^{\mathrm{cal}}(\mathbf{x})}\label{eq:image_phase_cal}
\end{equation}
with $\varphi_{nm}^{\mathrm{cal}}(\mathbf{x})$ denoting the space-varying calibration phase for the $nm$-th image, to be obtained with fine synchronization. 

Let us consider $P$ calibration points, where $P\leq Q$, located at $\{\mathbf{x}_p\}_{p=1}^P$ (true positions) and denote with $\delta \overline{\varphi}_{nm,p} = \angle I_{nm}\left(\mathbf{x}= \overline{\mathbf{x}}_{p}\right)$ the measured phase of the $p$-th target in the $nm$-th image. Here, $\overline{\mathbf{x}}_{p}$ is the location at which the target is observed in all the images, and it is reasonably close to the true position $\mathbf{x}_p$. A prevalent criterion employed for the selection of calibration points involves the identification of $P$ targets, consistently present in all the images, that exhibit the highest intensity in each image (permanent scatterers \cite{ferretti01}). 

Let us define the parametric model for the phase after the back-projection \eqref{eq:image_sync} with nominal parameters $\overline{\beta}_n$ and $\overline{\tau}_{nm,p} = \overline{\tau}_{nm}(\overline{\mathbf{x}}_p)$ as:
\begin{equation}\label{eq:phase_model_residual_abs}
\begin{split}
     \delta \varphi_{nm,p}  = &  + 2\pi f_{n} \left(1+ \beta_{n} \right) \tau_{nm,p} - \alpha_{nm} + \theta_p \\ & - 2\pi f_{n} \left(1+ \overline{\beta}_{n} \right) \overline{\tau}_{nm,p}.
\end{split}
\end{equation}
%
%
%
Since the target's phase $\theta_p$ is not of interest and it is common to all images, we enforce the difference with the phase of a reference image, e.g., the monostatic image used for coarse synchronization, such as
\begin{equation}\label{eq:phase_model_residual}
\begin{split}
    \Delta \varphi_{nm,p} &= \delta \varphi_{nm,p} - \delta \varphi_{\mathrm{ref},p}.
\end{split}
\end{equation}
The general fine phase synchronization problem can be formulated as
\begin{equation}\label{eq:phase_calibration_problem_formula}
    \widetilde{\boldsymbol{\xi}} = \underset{\boldsymbol{\xi} }{\text{argmax}} \,  \frac{1}{N^2P}\Re\left\{ \sum_{n,m,p} e^{j\left[\Delta\overline{\varphi}_{nm,p} -\Delta\varphi_{nm,p}(\boldsymbol{\xi}) \right]}\right\} 
\end{equation}
where $\widetilde{\boldsymbol{\xi}} = \left[\widetilde{\mathbf{s}}^T_1, ..., \widetilde{\mathbf{s}}^T_N,  \widetilde{\mathbf{x}}^T_1, ..., \widetilde{\mathbf{x}}^T_P, \widetilde{\alpha}_{1},...,\widetilde{\alpha}_N\right]^T$ denote the finely estimated positions of sensors and targets as well as clock offsets, respectively, while the term $\Delta\overline{\varphi}_{nm,p}=\delta\overline{\varphi}_{nm,p} - \delta\overline{\varphi}_{\mathrm{ref}}$ denotes the measured difference between $\delta\overline{\varphi}_{nm,p} = \angle I_{nm}\left( \overline{\mathbf{x}}_{p}\right)$ and $\delta\overline{\varphi}_{\mathrm{ref}} = \angle I_{\mathrm{ref}}\left(\overline{\mathbf{x}}_{p}\right)$. Problem \eqref{eq:phase_calibration_problem_formula} is not convex as it presents many local maxima that are difficult to handle, unless a-priori information is available \cite{Phase_cal_TGRS}. Therefore, we outline an alternating maximization algorithm in Section \ref{subsect:algorithm}, and we discuss the phase synchronization performance via HCRB in Section \ref{sect:HCRB}.

\subsection{Fine Synchronization Algorithm}\label{subsect:algorithm}

A practical solution to problem \eqref{eq:phase_calibration_problem_formula} is accomplished by an alternated maximization approach composed of 3 steps. 

\begin{enumerate}
    \item Estimation of residual sensors' positions, from monostatic images only, given the nominal targets' positions $\{\overline{\mathbf{x}}_p\}$
    \begin{equation}\label{eq:step1}
        \{\widetilde{\mathbf{s}}_n\}_{n=1}^N = \underset{ \{\mathbf{s}_n\}_{n=1}^N}{\text{argmax}}  \,\Re\left\{ \sum_{n,p} e^{j\left[\Delta\overline{\varphi}_{nn,p} -\Delta\varphi^{(1)}_{nn,p} \right]}\right\} 
    \end{equation}
    where $\Delta\varphi^{(1)}_{nn,p} = \Delta\varphi_{nn,p}(\mathbf{s}_n|\{\overline{\mathbf{x}}_p\} )$ is the differential parametric phase model described by \eqref{eq:phase_model_residual} conditioned to the nominal a-priori calibration points. Recall that by using monostatic images, it is $\alpha_{nm}=0$;
    
    \item Estimation of clock offsets $\{\alpha_{nm}\}$ from multistatic phase measurements according to model \eqref{eq:phase_model_residual}, given $\{\widetilde{\mathbf{s}}_n\}$ from step 1) and $\{\overline{\mathbf{x}}_p\}$, as
    \begin{equation}\label{eq:step2}
        \widetilde{\alpha}_{nm} = \frac{1}{P} \sum_p\left[\Delta \overline{\varphi}_{nm,p} - \Delta \varphi_{nm,p}(\widetilde{\mathbf{s}}_n, \widetilde{\mathbf{s}}_m, \overline{\mathbf{x}}_p)\right].
    \end{equation}
    \item Estimation of residual targets' positions from both monostatic and bistatic images using $\{\widetilde{\mathbf{s}}_n\}$ from step 1) and $\{\widetilde{\alpha}_{nm}\}$ from step 2)
    \begin{equation}\label{eq:step3}
        \{\widetilde{\mathbf{x}}_p\}_{p=1}^P = \underset{ \{\mathbf{x}_p\}_{p=1}^P}{\text{argmax}}  \, \Re\left\{ \sum_{n,m,p} \hspace{-0.1cm}e^{j\left[\Delta\overline{\varphi}_{nm,p} -\Delta\varphi^{(3)}_{nm,p} \right]}\right\} 
    \end{equation}
    where $\Delta\varphi^{(3)}_{nm,p} = \Delta\varphi_{nm,p}(\mathbf{x}_p|\widetilde{\mathbf{s}}_n,\widetilde{\mathbf{s}}_m,\widetilde{\alpha}_{nm})$ is the parametric phase model conditioned to the estimated sensors' positions and clock offsets.
\end{enumerate}
Optimization steps \eqref{eq:step1} and \eqref{eq:step3} can be tackled by gradient descent search over limited domains, approximately corresponding to a tenth of the resolution of single images, thanks to coarse synchronization \cite{Phase_cal_TGRS}. Such limited domains also guarantee that, in practical conditions, the optimization converges to a global optimum, for which no detrimental effect are observed on the combined image \eqref{eq:image_sync_tot}. The computation flow from 1) to 3) can be possibly iterated until a reasonable stopping threshold is matched; a direct possibility is to fix a minimum value of the cost function in \eqref{eq:phase_calibration_problem_formula} (that peaks at 1). After the fine synchronization procedure, the calibration phase is computed as
\begin{equation}\label{eq:calibration_phase}
    \varphi_{nm}^\mathrm{cal}(\mathbf{x}) = - 2 \pi f_n (1 \hspace{-0.05cm}+ \hspace{-0.05cm}\overline{\beta}_n) \left[\overline{\tau}_{nm}(\mathbf{x}) \hspace{-0.1cm}- \hspace{-0.1cm}\widetilde{\tau}_{nm}(\mathbf{x})\right] \hspace{-0.05cm}+ \hspace{-0.05cm}\widetilde{\alpha}_{nm}.
 \end{equation}

\subsection{Sufficient Conditions for Solving \eqref{eq:phase_calibration_problem_formula}}\label{subsect:inversion_analysis}

The inversion of \eqref{eq:phase_calibration_problem_formula} consists of the estimation the $3N + 2P$ parameters based on $\left(N^{2}-1\right)P$ observations (phase measurements).  A a sufficient condition for the inversion of \eqref{eq:phase_calibration_problem_formula}, that binds the number of selected points $P$ and the number of sensors $N$, can be derived from the inspection of the monostatic setup. Here, $\alpha_{nm}=0$ and the absolute frequency offset $\beta_n$ can be accurately retrieved up to a common term $\delta \beta$ (Section \ref{subsect:freq_est}). Based on \eqref{eq:phase_model_residual}, the special problem consists of estimating sensor and target positions based on the observed monostatic TOFs. In the 2D space, a fixed roto-translation of all the sensors and targets leaves TOFs unvaried, thus problem \eqref{eq:phase_calibration_problem_formula} in a monostatic setup can only be solved with no ambiguity by assuming that the absolute position of a sensor and at least one coordinate of one of the selected targets are known. In this setting, the number of phase observations is $\left(N-1\right)P$ while the number of unknowns is $2\left(N-1\right)+2P-1$. As a result, the sufficient condition to solve \eqref{eq:phase_calibration_problem_formula} is 
\begin{equation}\label{eq: special prob}
\left(N-1\right)P\geq2\left(N-1\right)+2P-1.
\end{equation}
It is worth noting that bistatic observations do not affect this evaluation. This is because bistatic TOFs are just linear combinations of monostatic TOFs, which means that they do not provide any new information about sensor and target positions. Additionally, terms $\{\alpha_{nm}\}$ in bistatic equations can be determined once the positioning is solved. Thus, problem \eqref{eq:phase_calibration_problem_formula} in a multistatic setup involves the estimation of $3\left(N-1\right) + 2P - 1$ parameters from $\left(N^2-1\right)P$ observations, and it can be solved up to a negligible global translation and rotation of the entire set of sensors and targets, as well as an indeterminate spatial contraction/dilation caused by the common frequency shift $\delta \beta$.

\begin{figure}[!b]
    \centering
    \subfloat[][$N=5$ sensors, $P=5$ calibration targets]{\includegraphics[width=\columnwidth]{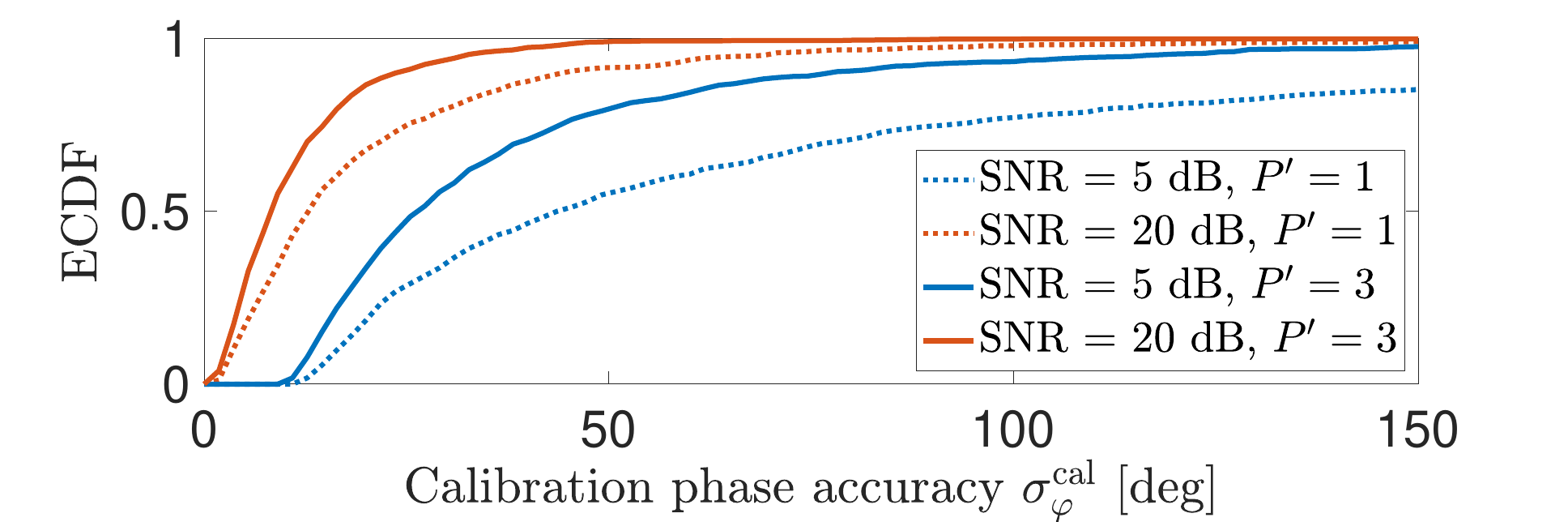}\label{subfig:5calibrationtargets}}\\
    \subfloat[][$N=5$ sensors, $P=10$ calibration targets]{\includegraphics[width=\columnwidth]{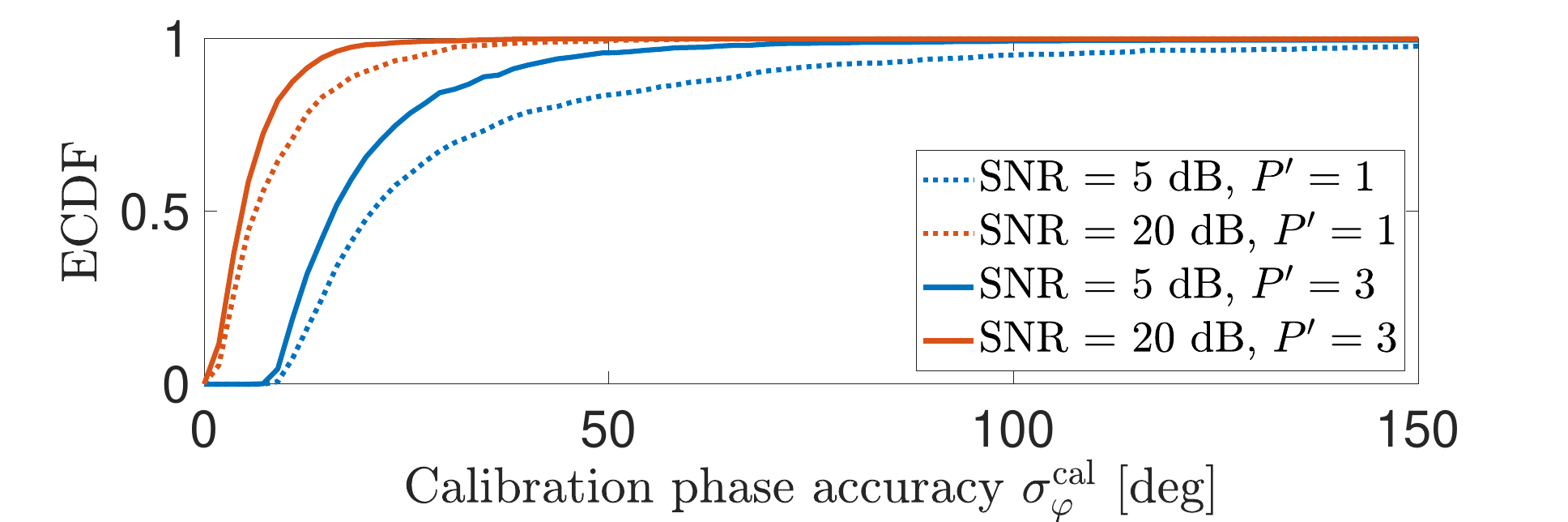}\label{subfig:10calibrationtargets}}\\
    \subfloat[][$N=3$ sensors, $P=P'=5$ calibration targets]{\includegraphics[width=\columnwidth]{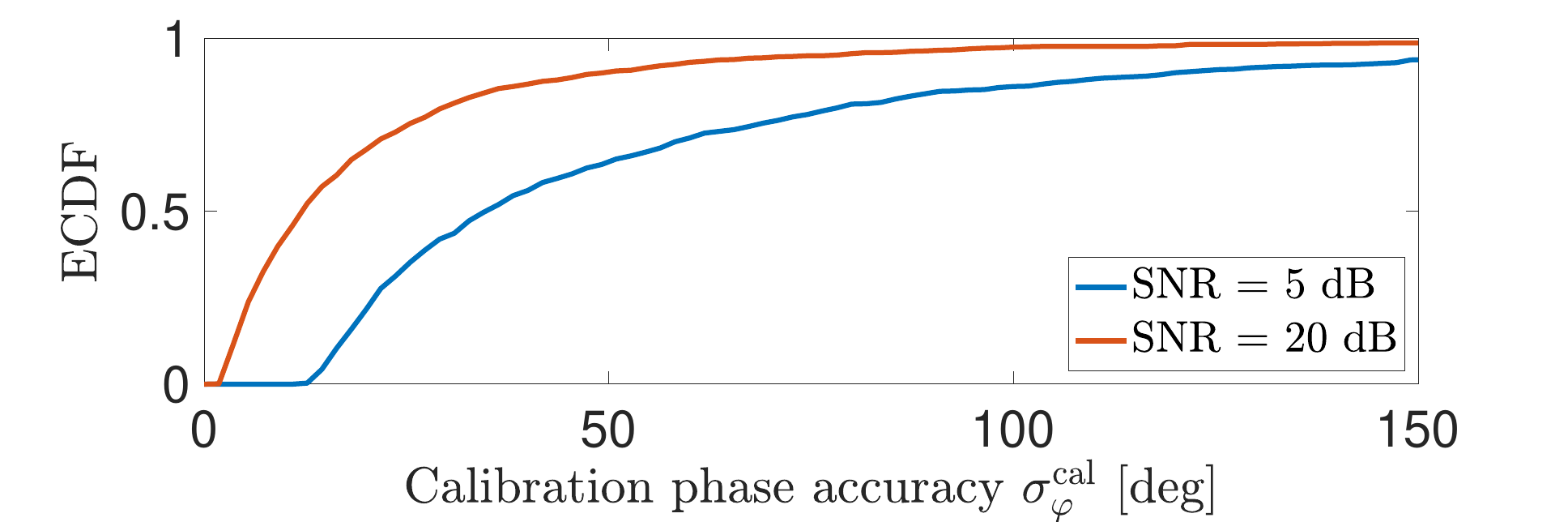}\label{fig_hcrb_3_sensors}}
    \caption{ECDF of phase calibration accuracy for (a) $N=5$ sensors with $P=5$ calibration targets (b) $N=5$ sensors with $P=10$ calibration targets and (c) $N=3$ sensors with $P=P'=5$ calibration targets, varying the SNR and the number of targets with prior knowledge $P'$. }
    \label{fig_hcrb}
\end{figure}

\section{Phase Synchronization Performance}\label{sect:HCRB}

Multistatic phase synchronization performance is here assessed by evaluating the empirical cumulative distribution function (ECDF) of the HCRB for the estimation of the $N^2-1$ calibration phases $\{\varphi_{nm}^\mathrm{cal}(\mathbf{x})\}$ in \eqref{eq:calibration_phase} for a selected point of interest, over random spatial distribution of vehicles in the scenario of Fig. \ref{fig:scenario}. Deterministic parameters are sensors' and calibration targets' positions $\{\mathbf{s}_n\}_{n=1}^{N-1}$ and $\{x_p\}_{p=1}^{P}$, $\{y_p\}_{p=1}^{P-1}$ as well as $\{\alpha_n\}_{n=1}^{N-1}$, in which $N-1$ accounts for the subtraction of the reference image in the phase measurement \eqref{eq:phase_model_residual} and we fix the $y$ coordinate of the $P$-th calibration target. When an a-priori knowledge about targets' positions is available, e.g., from previous sensing acquisitions, a subset $P'\leq P$ of the calibration targets can be treated as a random parameter with known covariance matrix.

The HCRB is evaluated by differentiating the phase differences $\Delta \varphi_{nm,p}$ modeled by \eqref{eq:phase_model_residual} w.r.t. to the parameters to be estimated. Following \cite{HCRB_07}, the HCRB on the estimation for deterministic parameters is obtained as:
\begin{equation}\label{eq:HCRB_1}
\mathbf{C}_d = \left(\mathbf{G}_{d}^T \left(\mathbf{C}_\varphi + \mathbf{G}_{r} \boldsymbol{\Xi}_r\mathbf{G}_{r}^T\right)^{-1}\mathbf{G}_{d}\right)^{-1}
\end{equation}
where \textit{(i)} $\mathbf{G}_{d}\in\mathbb{R}^{\left(N^{2}-1\right)P \times M_d }$ and $\mathbf{G}_{r}\in\mathbb{R}^{\left(N^{2}-1\right)P \times M_r }$ are the matrices built by taking the partial derivatives of the phase model \eqref{eq:phase_model_residual} w.r.t. the $M_d = 3(N-1) + 2(P-P')-1$ deterministic and $M_r = 2P'$ random parameters, respectively, \textit{(ii)} $\boldsymbol{\Xi}_r\in\mathbb{R}^{M_r \times M_r}$ is the a-priori available covariance matrix of random parameters, \textit{(iii)} $\mathbf{C}_\varphi = \sigma^2 (\mathbf{I} + \mathbf{1}\mathbf{1}^T)\in\mathbb{R}^{\left(N^{2}-1\right)P \times \left(N^{2}-1\right)P}$ is the covariance matrix of the phase noise. The latter is characterized by $\sigma^{2}\simeq 1/(2\,\mathrm{SNR})$, where the SNR is measured at the estimated location of each calibration target $\{\widetilde{\mathbf{x}}_p\}_{p=1}^P$ (here the SNR is considered approximately equal for all calibration targets, over all the $N^2-1$ images).
Notice that $\mathbf{G}_{r} \boldsymbol{\Xi}_r\mathbf{G}_{r}^T = \mathbf{C}^\mathrm{prior}_{\boldsymbol{\varphi}}$ is the covariance matrix of calibration phases due to a-priori information on parameters.
The covariance matrix of the calibration phases is obtained from \eqref{eq:HCRB_1} as:
\begin{equation}\label{eq: covariance cal}
\mathbf{C}^\mathrm{cal}_{\boldsymbol{\varphi}} = \mathbf{G}_{d,\mathrm{cal}} \mathbf{C}_{d,\mathrm{cal}} \mathbf{G}_{d,\mathrm{cal}}^T
\end{equation}
where $\mathbf{C}_{d,\mathrm{cal}}$ is obtained by taking the $3\left(N-1\right)\times3\left(N-1\right)$ block of $\mathbf{C}_{d}$ that corresponds to sensors' positions $\{\mathbf{s}_n\}_{n=1}^{N-1}$ and $\{\alpha_n\}_{n=1}^{N-1}$ , $\mathbf{G}_{d,\mathrm{cal}}$ is the $\left(N^{2}-1\right)\times3\left(N-1\right)$
matrix built by taking the partial derivatives of the phase model in \eqref{eq:phase_model_residual} w.r.t. sensors' positions and phase offsets.
Notice that, although the choice of the calibration targets $\{\mathbf{x}_p\}_{p=1}^P$ do affect the calibration phases via $\mathbf{C}_d$ in \eqref{eq:HCRB_1}, according to \eqref{eq:calibration_phase} the final calibration phases that affect the final image are only function of $\{\mathbf{s}_n\}_{n=1}^{N}$ and $\{\alpha_n\}_{n=1}^{N}$ (comprising the position and clock offset of the reference sensor).

To test the multistatic phase synchronization performance, we show the ECDF of the HCRB-derived calibration phase accuracy, defined as 
\begin{equation}\label{eq: rms error}
\sigma_{\varphi}^{\mathrm{cal}}=\sqrt{\frac{\mathrm{trace}\left(\mathbf{C}^\mathrm{cal}_{\boldsymbol{\varphi}}\right)}{N^{2}-1}}.
\end{equation}
for random vehicles' and calibration targets' positions $\{\mathbf{s}_n\}_{n=1}^N$ and $\{\mathbf{x}_p\}_{p=1}^P$, respectively. The reference geometry is the one in Fig. \ref{fig:scenario}. Vehicles' positions are varied uniformly within an interval of $50$ m along $x$ (along track), with a minimum distance of $7$ m between any two vehicles. Targets' positions are varied uniformly within an area of $40$ m along $x$ and $10$ m along $y$ (across-track). For each realization, we evaluate the HCRB on $\mathbf{C}^\mathrm{cal}_{\boldsymbol{\varphi}} $, assuming a prior confidence of $20$ cm along $x$ and $y$ about the positions of $P' < P$ targets (in $\boldsymbol{\Xi}_r$) for which we assume prior knowledge, which we deem to be realistic for the considered scenario.

The results are shown in Fig. \ref{fig_hcrb}. Figs. \ref{subfig:5calibrationtargets} and \ref{subfig:10calibrationtargets} show the CDF for $P=5$ and $P=10$ calibration targets (for $N=5$ vehicles and variable $P'$ and SNR). As expected, the phase synchronization performance improves by increasing both $P$ and $P'$. It is interesting to note the case with $P=10$ calibration targets with $P'=1$ prior target yields better performance than the case of $P=5$ calibration with $P'=3$ prior targets. This shows that, for the particular scenario herein considered, the availability of many targets in unknown positions is to be preferred to the case of prior knowledge about a few targets. When only $N=3$ sensors are available (Fig. \ref{fig_hcrb_3_sensors}), and we choose $P=P'=5$ calibration targets (i.e., we have a prior knowledge of all targets), we are in a condition for which \eqref{eq: special prob} is violated. Still, we observe $\sigma^{\mathrm{cal}}_{\varphi} \leq 50$ deg at SNR = $20$ dB in over $90\%$ of the tested geometrical configurations, thanks to the availability of a strong a-priori information about the position of calibration targets. Yet, it advocates for the feasibility of an accurate phase synchronization even in the case where the condition in \eqref{eq: special prob} is not fulfilled.

\section{Detection probability of a weak target}\label{sect:PCD}

The advantages of the cooperation among different sensors/vehicles can be demonstrated by evaluating the probability of correct detection (PCD) for a weak target close to a stronger one, such that the two targets are \textit{coupled} together in the overall image, as it is the case of a pedestrian crossing the road while in proximity to parked vehicles, as sketched in Figs. \ref{fig:scenario} and \ref{fig:compressive_sensing}.

\begin{figure}[!b]
    \centering
    \subfloat[][]{\includegraphics[width=0.9\columnwidth]{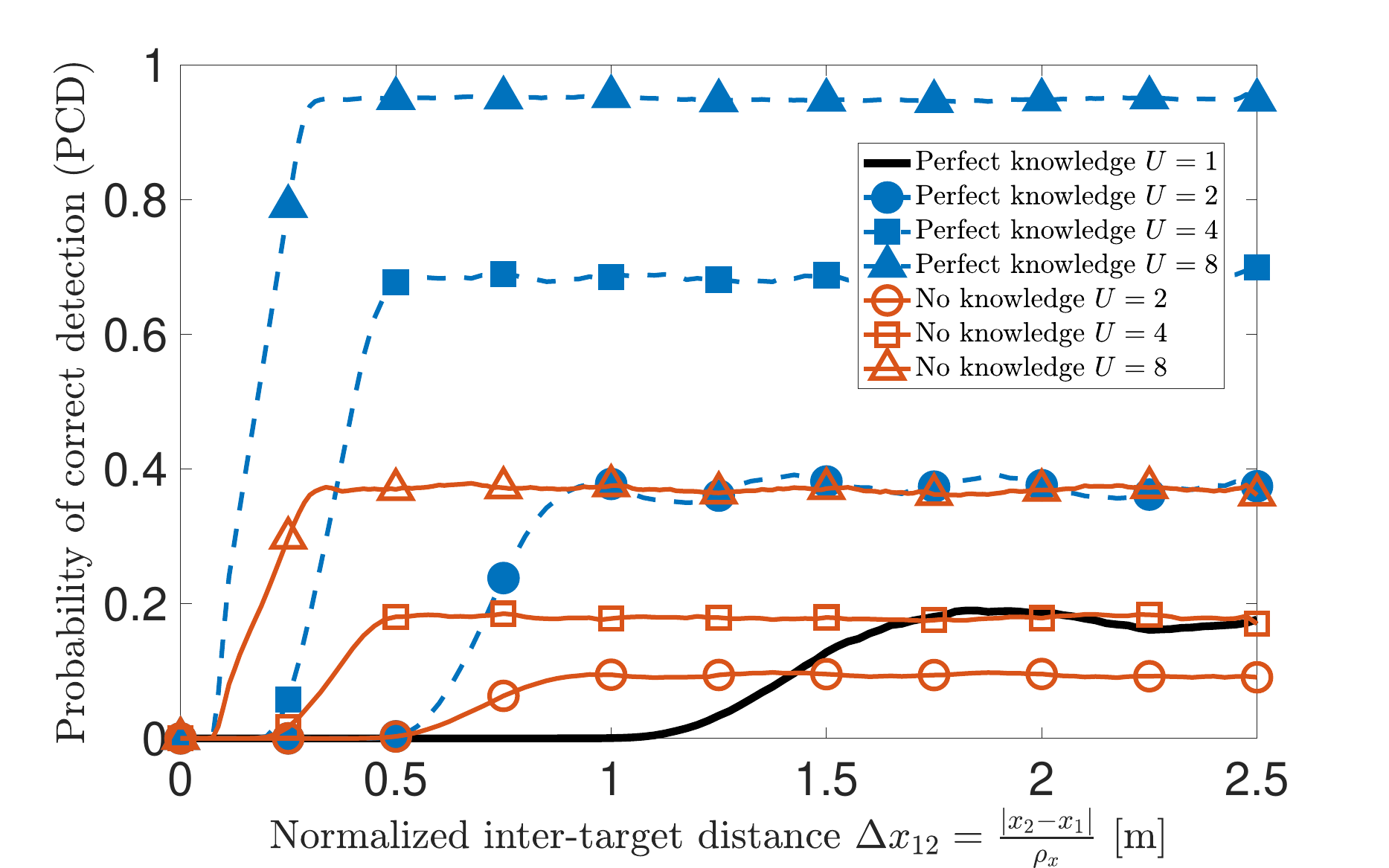}\label{eq:PCD_vs_Dx_noPhError}}\vspace{-0.25cm}\\
    \subfloat[][]{\includegraphics[width=0.9\columnwidth]{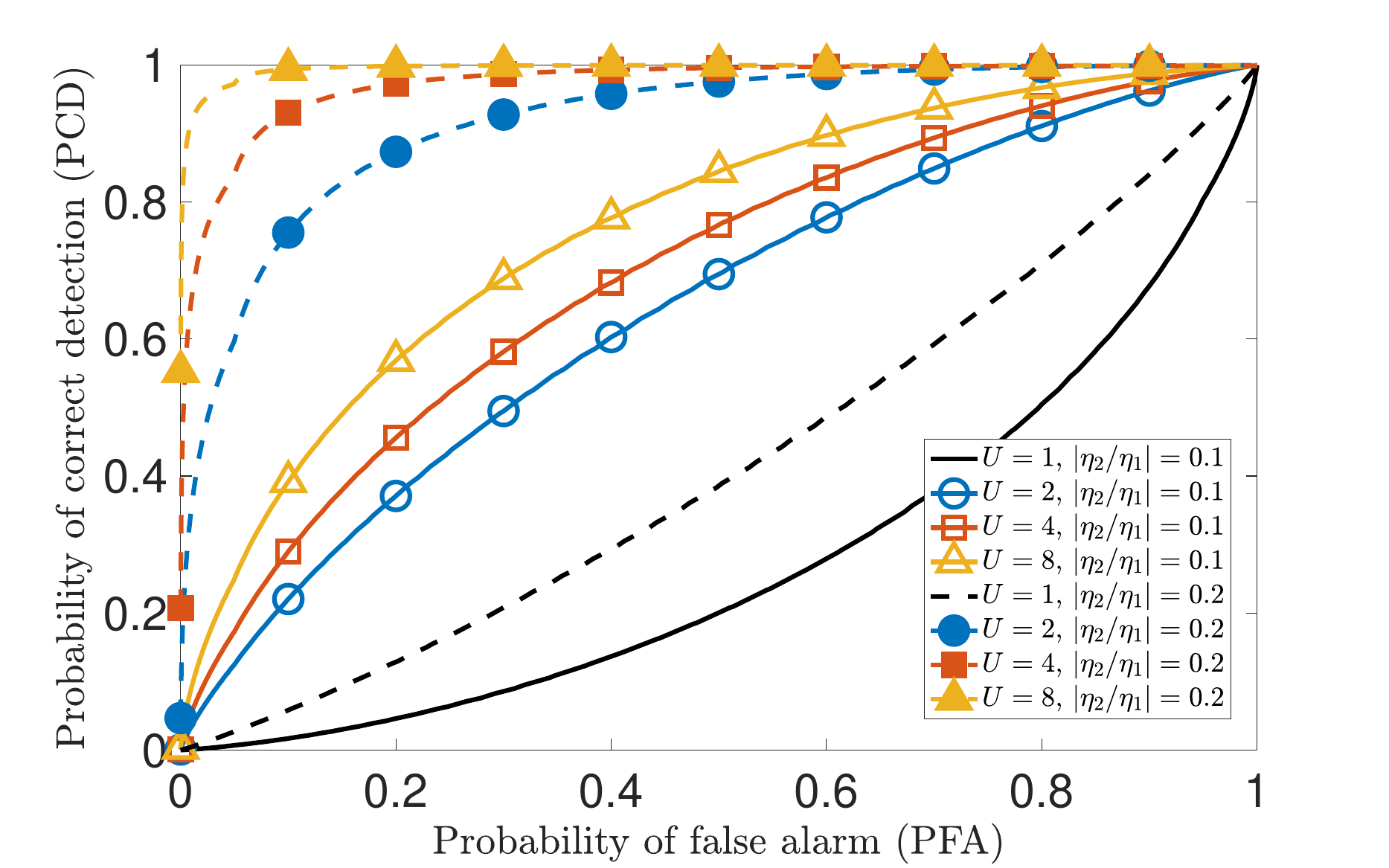}\label{eq:PCD_vs_PFA}}
    \caption{PCD vs. (a) inter-target distance for $|\eta_2/\eta_1|=0.1$ and (b) PFA (ROC) for $\Delta x_{12} = 1$ and $|\eta_2/\eta_1|=0.2$. The usage of multiple acquisitions increases both resolution and SNR, allowing the detection of a weak target in proximity to a strong one. }
    \label{fig:PCD_set1}
\end{figure}

Let us consider first the Rx signal model \eqref{eq:Rxsignal_wavenumbers} in the wavenumber domain for perfect synchronization. We can uniformly sample the 2D wavenumber domain to vectorize the observation
\begin{equation}\label{eq:Rxsignal_wavenumbers_discrete}
    \mathbf{y} = \mathbf{D}(\mathbf{x}_1,\mathbf{x}_2) \boldsymbol{\eta} + \mathbf{z}
\end{equation}
obtaining $N^2 |\mathcal{K}|$ measurements, being $|\mathcal{K}|$ the number of samples in the wavenumber domain (cardinality of the overall \textit{discrete} coverage set \eqref{eq:wavenumber_coverage_overall}), $\mathbf{D}(\mathbf{x}_1,\mathbf{x}_2)=[\mathbf{d}(\mathbf{x}_1), \mathbf{d}(\mathbf{x}_2)]\in \mathbb{C}^{N^2 |\mathcal{K}| \times 2}$, with $\mathbf{d}(\mathbf{x}_1) = e^{-j \mathbf{K}^T\mathbf{x}_1}$, $\mathbf{d}(\mathbf{x}_2) = e^{-j \mathbf{K}^T\mathbf{x}_2}$ is the model including targets' positions ($\mathbf{K}\in\mathbb{C}^{ 2 \times N^2 |\mathcal{K}|}$ is the matrix where the sampled wavenumbers are stacked in columns), $\boldsymbol{\eta}=[\eta_1,\eta_2]^T\in\mathbb{C}^{2 \times 1}$ denote the complex amplitudes of the two targets, including path-losses and phases $\theta_1$ and $\theta_2$. Noise in the wavenumber domain is such that $\mathbf{z} \sim \mathcal{CN}(\mathbf{0},\sigma_z^2\mathbf{I})$, where the power is $\sigma_z^2=N_0 \mathrm{d}k_x\mathrm{d}k_y$ ($\mathrm{d}k_x$ and $\mathrm{d}k_y$ are the sampling periods in the wavenumber domain). 
The problem of detecting a weak target near a strong one can be cast as a binary classification between two distinct hypotheses
\begin{equation}\label{eq:hypoteses_signal}
    \begin{dcases}
        \mathcal{H}_0: \mathbf{y} = \eta_1 \mathbf{d}(\mathbf{x}_1)  + \mathbf{z}\\
        \mathcal{H}_1: \mathbf{y} = \eta_1 \mathbf{d}(\mathbf{x}_1)  + \eta_2\mathbf{d}(\mathbf{x}_2) + \mathbf{z}
    \end{dcases}
\end{equation}
namely $\mathcal{H}_0$ represents the single target plus noise case and $\mathcal{H}_1$ two targets plus noise. 
For the detection, we consider the following scalar model of the observation \cite{ulaby82}
\begin{equation}\label{eq:model_PCD}
     v = \|\mathbf{y}\|^2 =\begin{dcases}
        \mathcal{H}_0: \|\eta_1\mathbf{d}(\mathbf{x}_1)  + \mathbf{z}\|^2\\
        \mathcal{H}_1: \|\eta_1 \mathbf{d}(\mathbf{x}_1)  + \eta_2\mathbf{d}(\mathbf{x}_2) + \mathbf{z}\|^2
    \end{dcases}
\end{equation} 
and we extend the generalized likelihood ratio test (GLRT) \cite{BigS} in two cases: \textit{(i)} the parameters of both targets, $\boldsymbol{\xi}_0 = [\mathbf{x}^T_1,\eta_1]^T$ (for $\mathcal{H}_0$) and $\boldsymbol{\xi}_1 = [\mathbf{x}^T_1,\mathbf{x}^T_2,\eta_1,\eta_2]^T$ (for $\mathcal{H}_1$) are known, and 
\textit{(ii)} the parameters of both targets are unknown and must be estimated prior to the GLRT.
In the first case (known $\mathbf{x}_1,\mathbf{x}_2,\eta_1,\eta_2$) the detection performance predicted by the GLRT is optimistic, as detecting the presence of a weak target given the a-priori knowledge of its position and amplitude (and also for the strong one) is relatively easy in moderate-to-high SNR regions. Still, this case represents a useful upper bound. Conversely, the second case (unknown $\mathbf{x}_1,\mathbf{x}_2,\eta_1,\eta_2$) provides a conservative prediction of the PCD, whereby the maximum likelihood (ML) estimation can be suboptimally approached by expectation-maximization \cite{BigS}, under Gaussian approximation for the distributions $p(v|\boldsymbol{\xi}_0,\mathcal{H}_0)$ and $p(v|\boldsymbol{\xi}_1,\mathcal{H}_1)$, which is reasonable under the central limit theorem for $2 N^2 |\mathcal{K}| > 100$. 

The PCD is analyzed in a 1D scenario, where the two targets are placed along the $x$ axis, and a variable number of acquisitions $U$ (an arbitrary combination of monostatic and bistatic images), each performed over a bandwidth $B=100$ MHz. We remark that the 1D case is presented for simplicity. The same conclusions apply to a generic 2D scenario with the due adaptations. In all the following, we apply the constant false alarm (CFAR) ratio approach to compute the PCD, considering a probability of false alarm (PFA) of $10^{-2}$. 

The first set of results are reported in Figure \ref{fig:PCD_set1}. Fig. \ref{eq:PCD_vs_Dx_noPhError} shows the PCD of the weak target varying the inter-distance from the strong one $\Delta x_{12} = |x_2-x_1|/\rho_x$ (normalized to the resolution of the single acquisition, $\rho_x = c/(2 B)\simeq 1.5$ m), for the following parameters: SNR for the strong target $|\eta_1|^2/\sigma_z^2 = 20$ dB, amplitude ratio $|\eta_2/\eta_1|=0.1$ (thus $|\eta_2|^2/\sigma_z^2 = 0$ dB), frequency sampling interval $\mathrm{d}f = 1$ MHz, number of acquisitions $U=1,4,8$. We consider that the central frequency $f_u$ of each acquisition is allocated such that to achieve a contiguous spectrum, $f_u = (u-1) B - UB/2$, for $u=1,...,U$, yielding the minimum level of sidelobes in the spatial waveform (see \cite{Tebaldini2017_tessellation}). Thus, the total bandwidth is $B_{tot}=400$ MHz (for $U=4$) and $B_{tot}=800$ MHz (for $U=8$). Fig. \ref{eq:PCD_vs_Dx_noPhError} demonstrates the benefit of coherent cooperation among multiple acquisitions: for fixed $\Delta x_{12}$, the PCD increases with $U$, as expected, due to the larger spectral coverage w.r.t. single acquisition, that brings a combined effect of improved resolution and increased SNR in the spatial domain. 
The PCD starts from 0 for $\Delta x_{12} = 0$, i.e., the weak target is superposed to the main lobe of the strong one and it is therefore not detectable and then increases up to a saturation value that jointly depends on the amplitude ratio $|\eta_2/\eta_1|$ and the SNR in the spatial domain SNR$_2$=$|\eta_2|^2 |\mathcal{K}|/\sigma_z^2$. For $U=1$, the PCD is substantially ruled by $|\eta_2/\eta_1|$ as SNR$_2$= 20 dB. The difference between knowing or not knowing the targets' parameters can be appreciated by comparing blue and red curves: the estimation of $(\mathbf{x}_1,\mathbf{x}_2,\eta_1,\eta_2)$ decreases the PCD but the improvement is consistent with $U$.  
Fig. \ref{eq:PCD_vs_PFA}, instead, shows the receiver operating characteristic (ROC) in the very same conditions of Fig. \ref{eq:PCD_vs_Dx_noPhError}, for a fixed inter-target distance $\Delta x_{12} = 1$ and $|\eta_2/\eta_1|=0.2$ (i.e., a slightly stronger second target) and considering the realistic case in $(\mathbf{x}_1,\mathbf{x}_2,\eta_1,\eta_2)$ are unknown and shall be estimated.

\begin{figure}[!b]
    \centering
    \includegraphics[width=0.9\columnwidth]{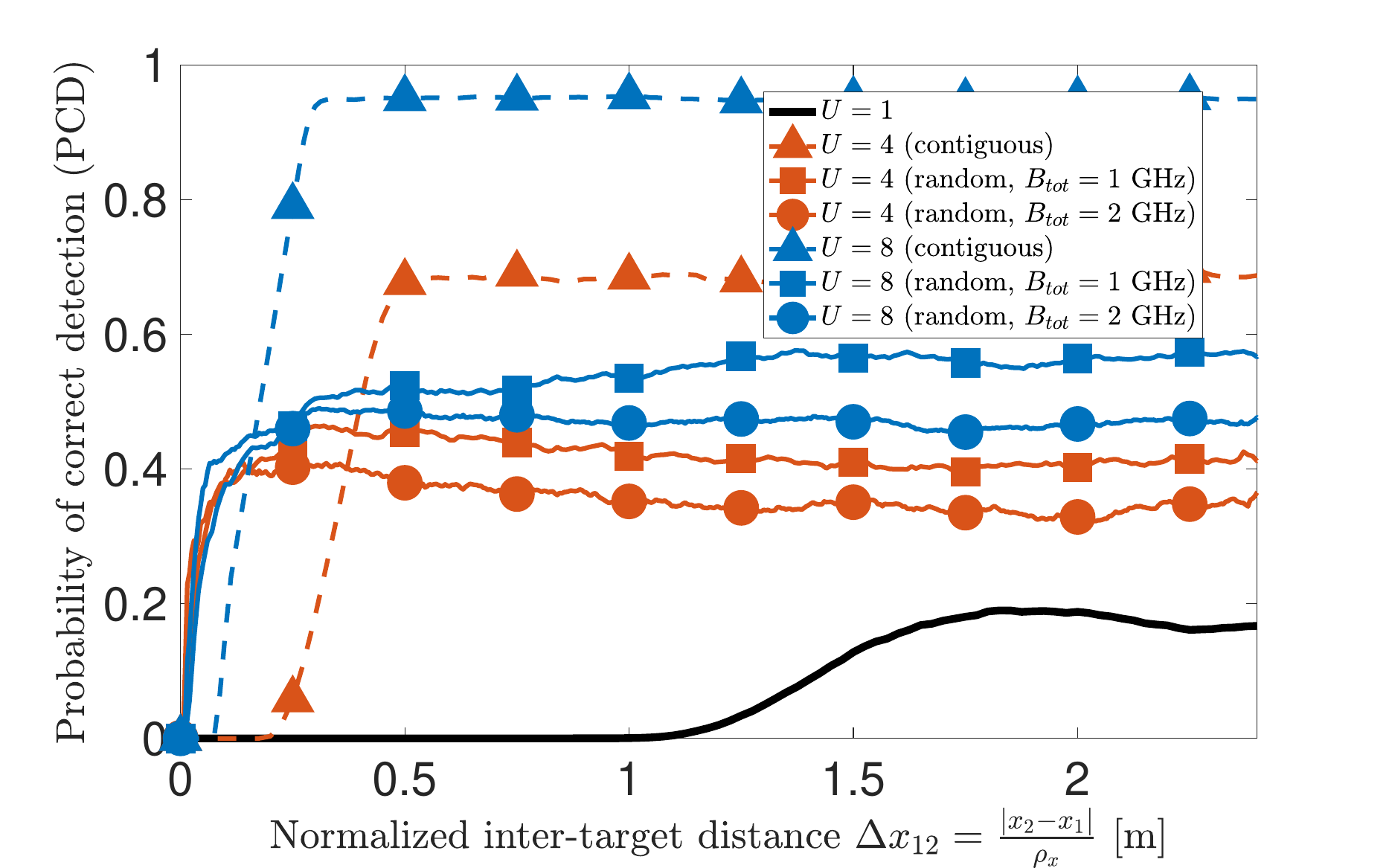}
    \caption{Impact of a non-contiguous bandwidth allocation on the PCD vs. inter-target distance. The PCD degrades w.r.t. to the case of a perfect contiguous allocation (wavenumber tessellation), still providing a remarkable advantage compared to the usage of a single sensor/acquisition.  }
    \label{fig:PCD_set2}
\end{figure}
\begin{figure}[!t]
    \centering
    \includegraphics[width=0.9\columnwidth]{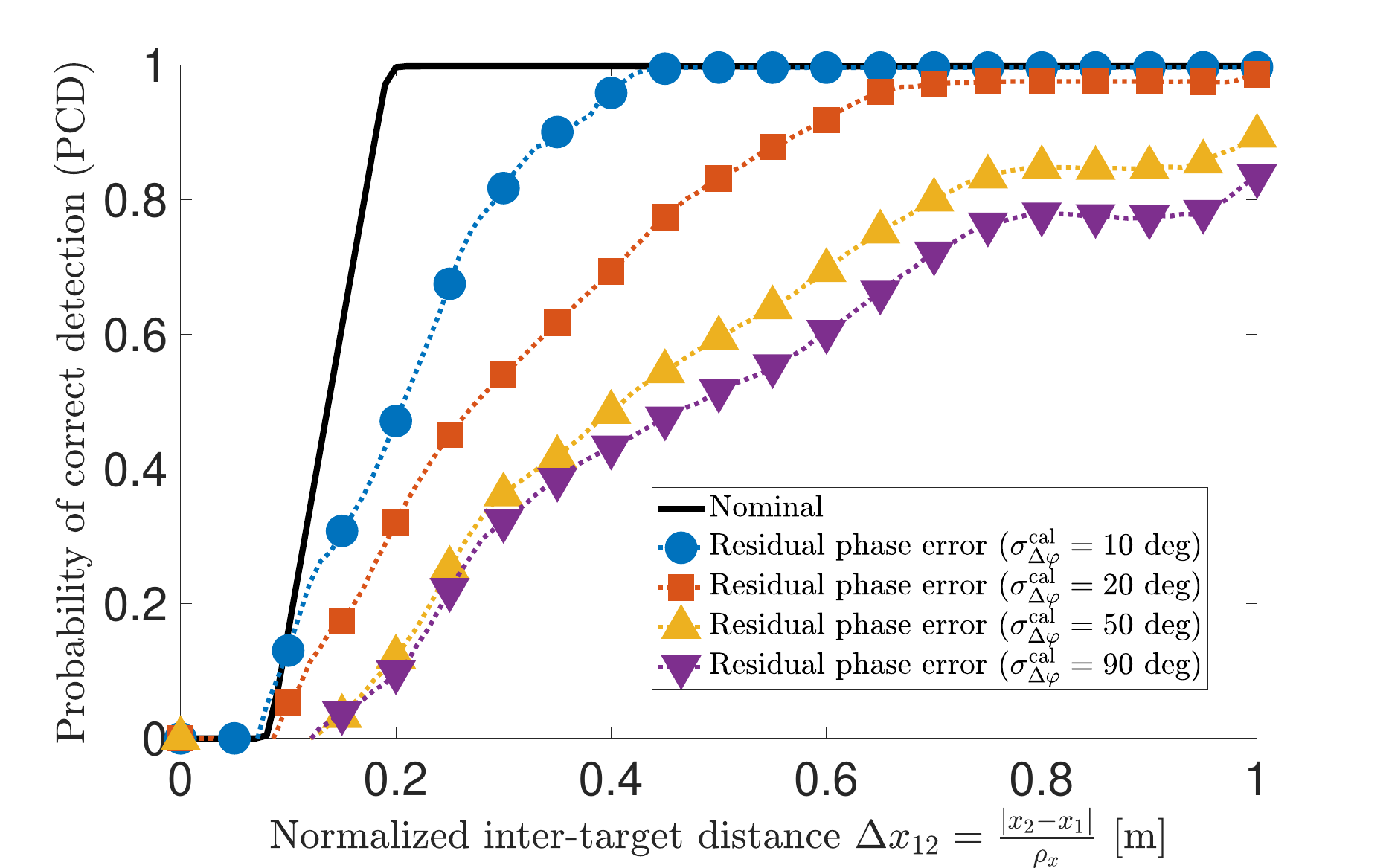}
    \caption{Impact of residual phase synchronization error on the PCD vs. inter-target distance for $U=8$ acquisitions. The PCD degrades proportionally to $\sigma_{\Delta \varphi}^\mathrm{cal}$, pushing for accurate phase synchronization before image fusion. }
    \label{fig:PCD_set3}
\end{figure}

When the single tiles in the wavenumber domain, i.e., the allocated bandwidths at each acquisition, are not contiguous---for example in a vehicular scenario, the wavenumber coverage is substantially ruled by vehicles' positions in space---, the average PCD performance degrades, still yielding substantial advantages w.r.t. the usage of a single sensor. Fig. \ref{fig:PCD_set2} shows the PCD for $U=4$ and $U=8$ acquisitions whose bandwidths are randomly and uniformly (still not overlapped) allocated over $B_{tot}=1$ GHz and $B_{tot}=2$ GHz, i.e., for increase spectrum sparsity, compared to the expected one for contiguous spectrum. As the average occupied bandwidth for random allocation is larger, the overall resolution increases (the PCD is $>0$ for decreasingly smaller values of $\Delta x_{12}$) but the PCD saturates to lower values, due to higher sidelobes. Remarkably, the improvement w.r.t. the single acquisition case is still significant. As a rule of thumb, the more the total bandwidth $B_{tot}$ for fixed $U$, the lower the performance, as the latter is inversely proportional to the sparsity spectrum index $U\times B/B_{tot}$. In any case, increasing $U$ yields less advantage compared to the ideal case of contiguous allocation. In this latter case, wavenumber tessellation and the resulting strategic resource allocation can be beneficial.

As a last result, we evaluate the impact of residual phase synchronization error, after the multistatic synchronization procedure of Section \ref{sect:_synchro_and_pos}. We consider as a benchmark the case in which the PCD is equal to 1 in the absence of residual phase synchronization errors (using $U=8$ acquisitions and for $|\eta_2/\eta_1|=0.2$). Then, we corrupt the observation model \eqref{eq:hypoteses_signal} with a random residual phase for each of the acquisitions, i.e., $\varphi_{n} \sim \mathcal{N}(0, (\sigma^{\mathrm{cal}}_{\varphi})^2)$, with $\sigma^{\mathrm{cal}}_{\varphi} = 10,20,50,90$ deg, leading to $\widetilde{\mathbf{d}}(x_1) = \mathbf{d}(x_1) \odot [e^{j \varphi_1}\mathbf{1}^T,...,e^{j  \varphi_N}\mathbf{1}^T]^T$ in \eqref{eq:hypoteses_signal}, where vector $\mathbf{1}$ has as many entries as the number of samples over the single bandwidth, i.e., $\lceil B/\mathrm{d}f \rceil $.  A phase error affects the waveform of the first target, whose sidelobes are, on average, higher compared to the ideal setup, hindering the correct detection of the weak target and mapping in an equivalent loss of resolution compared to the ideal setup (Fig. \ref{fig:PCD_set3}).  This highlights the importance of an accurate phase synchronization to boost the performance of a coherent sensing system.

\section{Experimental demonstration}\label{sec:real_data}

\begin{figure}[b]
\centering
\includegraphics[width=0.8\columnwidth]{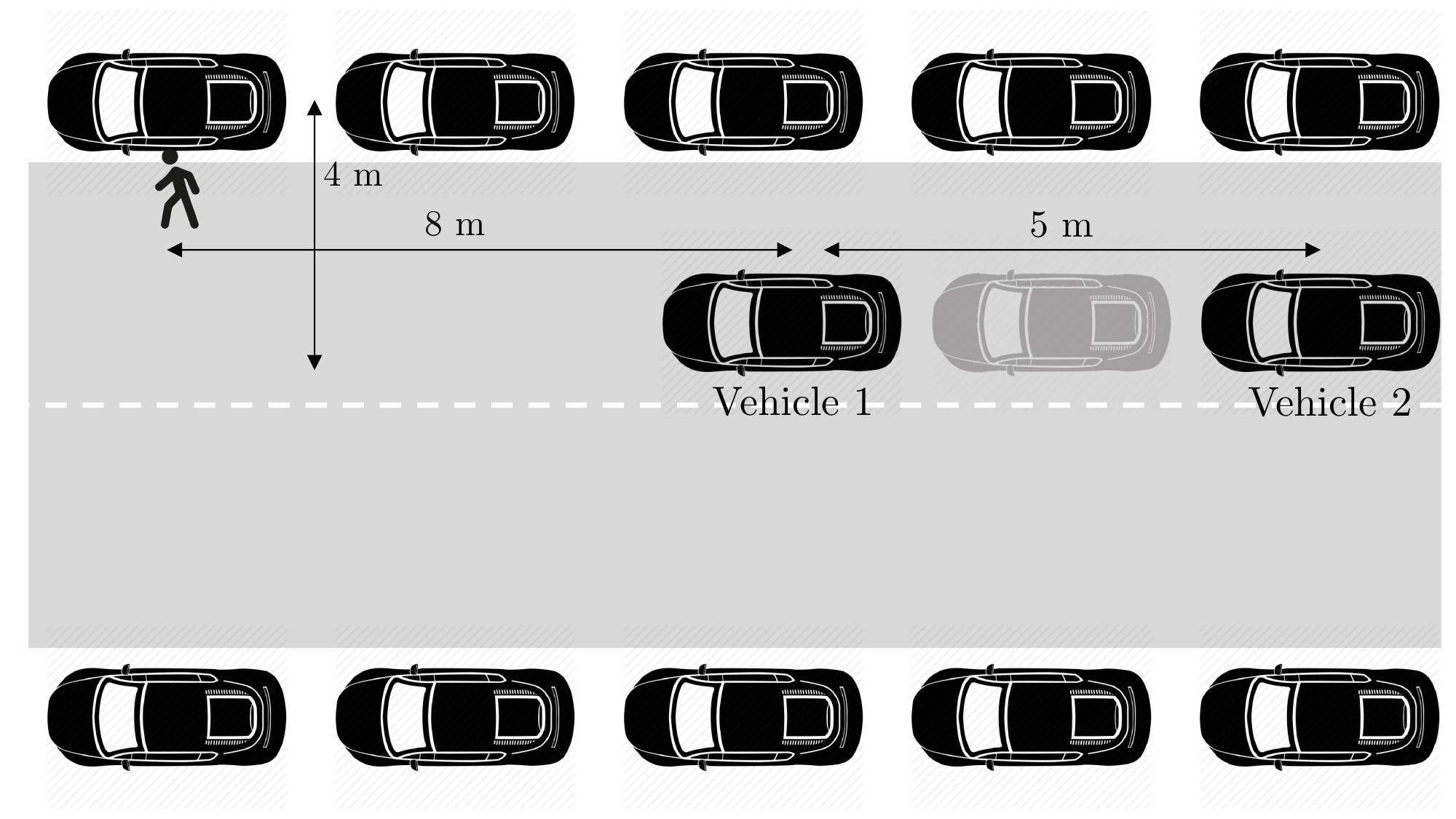}
\caption{Sketch of the experimental scenario used to assess the cooperation benefits in a real automotive networked sensing system. The grey car sketch represents the phase center of the emulated bistatic acquisition.}
\label{fig:scenario_reale}
\end{figure}

In this section, we further demonstrate the benefits of a coherent networked sensing system in a real vehicular scenario, presenting experimental data. In the considered setting, a moving vehicle is equipped with an 8-channel low-resolution front-looking radar operating at $f_0 = 78.5$ GHz over a bandwidth $B=180$ MHz, leading to a range resolution of $\simeq 83$ cm. To cater to automotive constraints \cite{noauthor_radiodetermination_2023}, the radar operates in \textit{burst} mode, alternating active and idle periods for a fixed duty cycle. For the full duration of each single burst, the moving vehicle synthesizes an aperture of $9$ cm, providing an azimuth resolution of $1.7$ deg at $45$ deg off-boresight, roughly corresponding to $60$ cm in cross-range at $20$ meters distance. These specifications correspond to recent off-the-shelf radars working in FDM, whose physical aperture is limited by cost \cite{TI_ref_MMWCAS}. 


\begin{figure}[!t]
\centering
\includegraphics[width=\columnwidth]{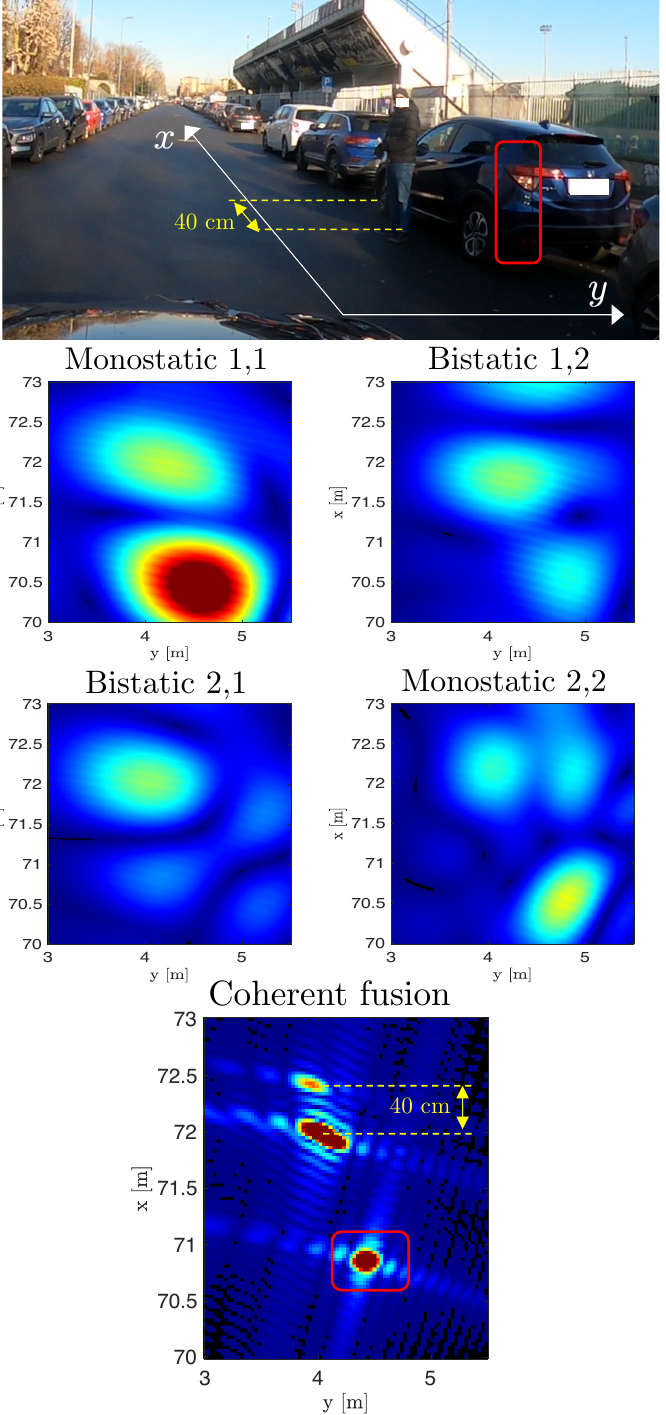}
\caption{Scenario (top), low-resolution monostatic and bistatic images (middle), high-resolution image (bottom). }
\label{fig:compressive_sensing}
\end{figure}

The sketch of the system geometry is depicted in Fig. \ref{fig:scenario_reale}. To emulate the desired networked sensing scenario, we exploit $3$ bursts of data, collected at $2.5$ meters distance from one to the other. This allows emulation of a real scenario in which two vehicles (vehicles 1 and 2) are platooning at a distance of $5$ m. The first and the third bursts of acquired data emulate monostatic acquisitions over two different bandwidths $f_1$ and $f_2$ ($|f_2-f_1| = B$), as for the FDM system described in the paper. The second burst "in between" the two cars emulate two bistatic measurements (1 to 2 and 2 to 1). Images from single bursts of data (both monostatic and bistatic) are first created exploiting SAR processing and on-board navigation sensors \cite{Tagliaferri2021_SARnavigation}. Then, single images underwent to the multistatic synchronization procedure outlined in Section \ref{sect:_synchro_and_pos}. 
The photo of the considered scenario is in Fig. \ref{fig:compressive_sensing} (top). A pedestrian is standing still in the roadway next to a parked car. This is one of the most challenging scenarios for automotive radar systems, as when a comparably weaker target (the pedestrian) is close to a stronger target (the car), the latter usually hinders the detection of the former as far as a low-resolution sensing system is employed. Indeed, a very high-resolution system is needed to distinguish and detect both targets. Fig. \ref{fig:compressive_sensing} (middle) shows single images of the area under test (car+pedestrian, monostatic and bistatic), where the pedestrian cannot be clearly discerned due to the low-resolution.  

Nonetheless, the cooperation improves the performance. A simple coherent sum of the images would lead to a high-resolution image, but very hard to interpret due to excessive amount of sidelobes. To this end, compressive sensing \cite{hadi_compressive_2015} can be exploited to reconstruct the scene with a lower level of sidelobes, provided that an initial estimate of the number of targets (possibly inferred by low-resolution images) is available \cite{rs15030845}. Fig. \ref{fig:compressive_sensing} (bottom) shows the coherent combination of both the monostatic and the bistatic images of the scene, after the application of compressive sensing. In the red box, the parked car results as a strong target. Notice that only the rear left corner of the parked car is visible, as the car doors act as specular mirrors for the incident signal. Remarkably, the legs of the pedestrian at 40 cm distance one to the other are now distinguishable as two bright spots. We remark that detecting people's legs has been demonstrated in \cite{rs14153602} for high-resolution radar imaging, exploiting GHz-wide bandwidths and long synthetic apertures, but this is one of the first experimental demonstrations under low-resolution constraints and within a network scenario. This experiment proves the validity and the benefits of networked sensing for automotive systems, showing that even though the single radar resources are insufficient to detect a weak target, the unleashing of the true sensing potential can be achieved through a coherent network sensing at the physical acquisition level.

\section{Conclusions}\label{sec:conclusions}
This study delves into the realm of coherent networked radio imaging, showcasing its significant benefits in a vehicular context and discussing synchronization challenges. The proposed use case is the VRU protection in a road safety context, where multiple vehicles aim at detecting a weak target, e.g., a pedestrian, in close proximity of a comparably stronger one, e.g., a parked car or a building, to take proper driving actions. Each vehicle is equipped with low-end off-the-shelf hardware, operating in FDM and multistatic fashion, and it is unable to distinguish the two targets unless a proper cooperation in the vehicular network is enforced to enable a coherent multistatic image fusion. We substantiate the effectiveness of coherent networked imaging through both theoretical analysis, which includes deriving the PCD of the weak target conditioned to the presence of the stronger one, and practical experimentation utilizing real data from an automotive radar in open road conditions. Remarkably, experimental results confirm that coherent cooperation among vehicles leads to a substantial performance improvement compared to the single sensor case, allowing to observe the legs of a pedestrian in close proximity of a parked car by coherently fusing data of two vehicles. Moreover, the work quantitatively discusses the tight requirements on the multistatic phase synchronization, i.e., clock synchronization plus sensors' positioning, that is a necessary enabler for the creation of high-quality environmental images. We also present a general cooperative multistatic synchronization procedure, herein applied to the specific vehicular scenario under consideration.
We deem future research should focus on refining cooperation methods concerning latency and detailing the required signaling and data exchange, as well as on the development of algorithms for real time processing. Moreover, we deem the considered vehicular case study should be further investigated by analysis of experimental data acquired in a wide range of conditions. 



\appendices

\appendices

\section{Background on diffraction tomography}
\label{sect:FEDT}

We here recall the basics of diffraction tomography, that allows quantifying the resolution of a radio imaging experiment from inspection of its spectral coverage \cite{WuToksoz1987}. Consider a 2D bistatic radio sensing experiment where a Tx $\mathbf{s}_\mathrm{Tx}$ and a Rx in $\mathbf{s}_\mathrm{Rx}$ illuminate a target in $\mathbf{x}_t$ with a plane wave at frequency $f_0$. Omitting geometrical energy losses and noise, which are irrelevant for this description, the Tx signal is
\begin{equation}\label{eq:Rxsignal_exttarget_simple}
\begin{split}
y(\mathbf{s}_\mathrm{Tx},\mathbf{s}_\mathrm{Rx},&\mathbf{x}_t)  \approx e^{-j k_0 (D_\mathrm{Tx}(\mathbf{x}_t) +  D_\mathrm{Rx}(\mathbf{x}_t))} \times \\ 
& \underbrace{ \iint_S \gamma(\mathbf{x}) e^{-j k_0 \left[\nabla D_{\mathrm{Tx}}(\mathbf{x}_t) + \nabla D_{\mathrm{Rx}}(\mathbf{x}_t)\right]^T \mathbf{x}}\;\mathrm{d}\mathbf{x}}_{\Gamma(\mathbf{k} =\mathbf{k}_{\mathrm{Tx}}-\mathbf{k}_{\mathrm{Rx}})}.
\end{split}
\end{equation}
under Born-Oppenheimer weak scattering condition (see \cite{Wolf1969}), the integral spans the visible area $S$ of the target in $\mathbf{x}_t$, $k_0 = 2\pi f_0/c$ is the fundamental wavenumber, $D_{\mathrm{Tx}}(\mathbf{x}_t) = \|\mathbf{x}_t - \mathbf{s}_\mathrm{Tx}\|$ and $D_{\mathrm{Rx}}(\mathbf{x}_t)=\|\mathbf{s}_\mathrm{Rx}-\mathbf{x}_t\|$, $\gamma(\mathbf{x})\in\mathbb{C}$ is complex reflectivity map of the environment, that for a point target it is $\gamma(\mathbf{x}) = \gamma_t \delta(\mathbf{x}-\mathbf{x}_t)$, and
\begin{equation}
    \nabla D_\mathrm{Tx}(\mathbf{x}_t) = \frac{\mathbf{x}_t - \mathbf{s}_\mathrm{Tx}}{\|\mathbf{x}_t - \mathbf{s}_\mathrm{Tx}\|} ,\,\, \nabla D_{\mathrm{Rx}}(\mathbf{x}_t) =\frac{\mathbf{s}_\mathrm{Rx} - \mathbf{x}_t}{\|\mathbf{s}_\mathrm{Rx} - \mathbf{x}_t\|}.
\end{equation}
Term $\Gamma(\mathbf{k})$ denotes the Fourier transform (FT) of the target's reflectivity $\gamma(\mathbf{x})$ evaluated in the \textit{wavenumber} $\mathbf{k} = \mathbf{k}_{\mathrm{Tx}}-\mathbf{k}_{\mathrm{Rx}}$\footnote{Wavenumbers are herein intended as linear combinations of wavevectors, i.e., 2D spatial frequencies.}, whereby oriented vectors
\begin{equation}\label{eq:wavevectors}
    \begin{split}
        \mathbf{k}_{\mathrm{Tx}}(\mathbf{s}_\mathrm{Tx},\mathbf{x}_t) & = (2 \pi f_0/c) \nabla D_{\mathrm{Tx}}(\mathbf{x}_t) \\
        \mathbf{k}_{\mathrm{Rx}}(\mathbf{s}_\mathrm{Rx},\mathbf{x}_t) & = - (2 \pi f_0/c) \nabla D_{\mathrm{Rx}}(\mathbf{x}_t)
    \end{split}
\end{equation}
denote plane wavevectors from Tx to the target and from the target to Rx, respectively. Any radio sensing measurement, indeed, simply \textit{excites} some target's wavenumbers $\mathbf{k}$. When the relative position between the sensor and the target changes, different wavenumbers are excited leading to a different resolution. 
For instance, if Tx employs a pulse with finite bandwidth $B$, and both Tx and Rx implement spatial \textit{apertures} $\mathcal{A}_\mathrm{Tx}$ and $\mathcal{A}_\mathrm{Rx}$ (i.e., using physical antenna arrays or synhtetic ones) the set of observed wavenumbers is the 2D region in $(k_x,k_y)$:
\begin{equation}\label{eq:wavenumber_coverage_two_arrays}
    \mathcal{K}_{\mathrm{Tx},\mathrm{Rx}}(\mathbf{x}_t) = \bigcup_{\substack{\mathbf{s}_\mathrm{Tx}\in\mathcal{A}_\mathrm{Tx}\\\mathbf{s}_\mathrm{Rx}\in\mathcal{A}_\mathrm{Rx}}}  \mathcal{K}_{B}(\mathbf{s}_\mathrm{Tx}, \mathbf{s}_\mathrm{Rx}, \mathbf{x}_t)
\end{equation}
where $\mathcal{K}_{B}(\mathbf{s}_\mathrm{Tx}, \mathbf{s}_\mathrm{Rx}, \mathbf{x}_t)$ is the wavenumber coverage of a single Tx-Rx antenna pair
\begin{equation}\label{eq:wavenumber_region_segment}
    \mathcal{K}_{B}(\mathbf{s}_\mathrm{Tx}, \mathbf{s}_\mathrm{Rx}, \mathbf{x}_t) = \left\{ \mathbf{k}(f) \, \bigg\lvert \, f \in \left[-\frac{B}{2}, +\frac{B}{2}\right]\right\}
\end{equation}
in which $\mathbf{k}(f) = \mathbf{k}_{\mathrm{Tx}}(\mathbf{s}_\mathrm{Tx}, \mathbf{x}_t, f)- \mathbf{k}_{\mathrm{Rx}}(\mathbf{s}_\mathrm{Rx}, \mathbf{x}_t, f)$ is frequency-dependent wavevector obtained by plugging $f_0+f$ in place of $f_0$ in \eqref{eq:wavevectors}. 

The sensing experiment resolution at a scattering point $\mathbf{x}_t$ in Cartesian coordinates can be directly quantified from the wavenumber coverage as follows 
\begin{equation}\label{eq:2Dresolution}
     \rho_x(\mathbf{x}_t)\approx \frac{2\pi}{\Delta k_x},\,\,\,\rho_y(\mathbf{x}_t)\approx \frac{2\pi}{\Delta k_y},
\end{equation}
where $\Delta k_x$ and $\Delta k_y$ are the widths of the wavenumber spectral coverage $\mathcal{K}_{\mathrm{Tx},\mathrm{Rx}}(\mathbf{x}_t)$ along $x$ and $y$, respectively. As far as more sensors are involved, with a disjoint set of covered wavenumbers, $\Delta k_{x,tot}$ and $\Delta k_{y,tot}$ are defined over the \textit{union} of covered regions. However, if the resulting spectrum is sparse, sideblobes in the final image appear. 

\bibliographystyle{IEEEtran}
\bibliography{Biblio/Bibliography,Biblio/insarnew,Biblio/insarref,Biblio/Unificata}

\end{document}